\pdfoutput=1
\documentclass[11pt,twoside,a4paper,cmspaper,final,collab]{cms-tdr}
\def\svnVersion{b9d0cfe}\def\svnDate{2023/09/11}\def\cmsCernNoTag{CERN-EP-2022-268}\def\cmsCernDate{\today}\def\cmsMessage{Published in the Journal of High Energy Physics as \href{http://dx.doi.org/10.1007/JHEP09(2023)051}{\doi{10.1007/JHEP09(2023)051}.}}
\begin{document}\cmsNoteHeader{EXO-21-009}

\newcommand{\WpCoupling}{\ensuremath{g_{\PWpr}}\xspace}
\newcommand{\Wcoupling}{\ensuremath{g_{\PW}}\xspace}

\cmsNoteHeader{EXO-21-009}
\title{Search for new physics in the \texorpdfstring{\PGt}{tau} lepton plus missing transverse momentum final state in proton-proton 
collisions at \texorpdfstring{$\sqrt{s} = 13\TeV$}{sqrt(s) = 13 TeV}}

\date{\today}

\abstract{A search for physics beyond the standard model (SM) in the final state with a hadronically decaying tau lepton and a neutrino is presented. This analysis is based on data recorded by the CMS experiment from proton-proton collisions at a center-of-mass energy of 13\TeV at the LHC, corresponding to a total integrated luminosity of 138\fbinv. The transverse mass spectrum is analyzed for the presence of new physics. No significant deviation from the SM prediction is observed. Limits are set on the production cross section of a \PWpr boson decaying into a tau lepton and a neutrino. Lower limits are set on the mass of the sequential SM-like heavy charged vector boson and the mass of a quantum black hole. Upper limits are placed on the couplings of a new boson to the SM fermions. Constraints are put on a nonuniversal gauge interaction model and an effective field theory model. For the first time, upper limits on the cross section of $t$-channel leptoquark (LQ) exchange are presented. These limits are translated into exclusion limits on the LQ mass and on its coupling in the $t$-channel. The sensitivity of this analysis extends into the parameter space of LQ models that attempt to explain the anomalies observed in \PB meson decays. The limits presented for the various interpretations are the most stringent to date. Additionally, a model-independent limit is provided.}

\hypersetup{
pdfauthor={CMS Collaboration},
pdftitle={Search for new physics in the tau plus missing transverse momentum final state in proton-proton collisions at sqrt(s)= 13 TeV},
pdfsubject={CMS},
pdfkeywords={CMS, exotica, tau, leptoquarks, Wprime, QBH}}

\maketitle 

\section{Introduction}
\label{sec:introduction}

A search for new phenomena is conducted in the final state with a
high transverse-momentum (\pt) tau lepton (\PGt) and large missing transverse momentum
due to the neutrino (\PGn). 
This search uses proton-proton 
($\Pp\Pp$) collision data at $\sqrt{s}=13\TeV$ recorded by the CMS experiment at the CERN LHC  
in 2016--2018, corresponding to an integrated luminosity of 138\fbinv.

Due to its short lifetime, the \PGt lepton itself is not detected 
directly, but 
via its decay products. The hadronic decays of the \PGt
lepton constitute about 2/3 of all its decays.
The hadronic decay products, 
\tauh, of the tau lepton are distinctive because they result in jets with a lower charged-hadron multiplicity
compared to the jets originating from the hadronization of partons produced in the hard scattering process. 
The selection of the hadronic \PGt decay mode also allows for discrimination against the leptonic \PGt decays,
which cannot be distinguished from prompt 
decays to light leptons, such as those examined in the most recent searches by the ATLAS and CMS Collaborations for 
\PWpr in the final states with an electron or muon plus a 
neutrino~\cite{ATLAS:2019lsy, CMS:2022yjm}.

The transverse mass, \mT, of the \PGt-\ptmiss system is the primary discriminating variable in this analysis.
It is defined as
\begin{equation}
 \mT = \sqrt{2\pt^{\tauh}\ptmiss [1-\cos{\Delta\phi(\vec{p}_{\text{T}}^{\tauh}, \ptvecmiss)}]},
\label{equ:mt}
\end{equation}
where $\pt^{\tauh}$ is the magnitude of the transverse momentum of the reconstructed \tauh candidate, 
\ptvecmiss is the vector of transverse momentum misbalance in the event,
\ptmiss is its magnitude, and 
$\Delta\phi(\vec{p}_{\text{T}}^{\tauh}, \ptvecmiss)$ is the azimuthal opening angle between the directions of 
\ptvecmiss and the tau lepton.  
We perform the search in a kinematic region where the $\vec{p}_{\text{T}}^{\tauh}$
and the \ptvecmiss are 
approximately balanced and back-to-back in the azimuthal plane.

A new charged gauge boson (\PWpr) is one possible signature of the sequential SM 
(SSM)~\cite{Altarelli:1989ff}. In this model, the \PWpr boson couplings to fermions are similar to those 
of the SM W, while the
additional decay mode of the \PWpr boson to top and bottom quarks is allowed for a sufficiently high boson mass. 
Searches for an SSM-like 
\PWpr boson decaying to the \PGt+$\PGn_{\PGt}$ final state have previously been reported by the 
CMS 
Collaboration at 8\TeV~\cite{Khachatryan:2015pua} and 13\TeV~\cite{Sirunyan_2019}, as well as by the 
ATLAS Collaboration at 13\TeV~\cite{Aaboud:2018vgh}. 
These searches by the ATLAS and CMS experiments excluded \PWpr bosons with masses up to 3.7\TeV and 4.0\TeV, respectively.
The \PWpr coupling strengths may differ from that of the SM \PW or the SSM \PWpr bosons. 
Therefore, we also consider the case where the ratio $\WpCoupling/\Wcoupling$ of the \PWpr boson 
coupling strength to that 
of the SM \PW boson is not unity.
The nonuniversal gauge interaction model (NUGIM) 
predicts an enhanced branching fraction for decays into third-generation fermions.
We also place upper exclusion limits on the ratio of the \PWpr boson coupling strength to the \PW boson 
coupling strength and on the NUGIM.

Some models of leptoquarks (LQs) are particularly interesting  in light of
the significant deviations from
the SM predictions in measurements in the flavor sector
of the $\PB\to\PD^{(*)} \Pell \PGn$ 
transitions,
by the BaBar~\cite{Lees:2012xj, Lees:2013uzd}, 
Belle~\cite{Huschle:2015rga, Hirose:2016wfn, Hirose:2017dxl, Belle:2019rba}, 
and LHCb~\cite{Aaij:2015yra, Aaij:2017uff, 
Aaij:2017deq, LHCb:2023zxo} Collaborations.
Separate measurements of the
$\PB\to \PK^{(*)} \Pell^{+}\Pell^{-}$
transitions by the LHCb Collaboration~\cite{Aaij:2019wad, Aaij:2017vbb, LHCb:2021trn, LHCb:2021lvy},
which previously found 
deviations, have recently been superseded by an updated result that is compatible with the SM 
predictions within 
uncertainties~\cite{LHCb:2022zom}.
The postulate of \TeV-scale LQs with dominant couplings to third-generation fermions is among the favored 
theoretical attempts to explain these anomalies~\cite{Tanaka:2012nw, Sakaki:2013bfa, Dorsner:2013tla, Dumont:2016xpj, 
Dorsner:2017ufx, Becirevic:2018afm, Crivellin:2017zlb, Gripaios:2014tna, Bauer:2015knc, ColuccioLeskow:2016dox,
Becirevic:2017jtw,Hiller:2017bzc, Hiller:2018wbv, DiLuzio:2017vat, DiLuzio:2018zxy, Calibbi:2017qbu, Bordone:2017bld, 
Barbieri:2017tuq, Buttazzo:2017ixm, Marzocca:2018wcf, Baker:2019sli, Cornella:2019hct, Cornella:2021sby}.
While numerous searches for single or pair production of LQs with dominant couplings to third-generation fermions at 
the LHC have yielded null results~\cite{CMS:2018qqq, ATLAS:2020dsf, ATLAS:2021yij, CMS:2020wzx, ATLAS:2020xov, 
CMS:2018oaj, ATLAS:2021oiz, CMS:2018svy, ATLAS:2021jyv, CMS:2018iye, CMS:2018txo}, the first search for LQ 
$t$-channel exchange in the \PGt+\PGn final state is presented here.

If the excess seen in the $\PB\to\PDs$ transitions is indeed due to new physics, the mass scale of its 
manifestation is expected to be at the TeV 
scale or not far beyond~\cite{DiLuzio:2017chi}.
To provide a different test for the presence of new physics, a description based on an effective 
field theory (EFT)~\cite{Greljo:2018tzh} with different 
types of coupling coefficients referred to as Wilson coefficients is also considered.

An open issue in modern particle physics is the large relative difference between the energy scale of 
electroweak symmetry breaking and that of gravity. This large discrepancy is often referred to as the hierarchy 
problem~\cite{PhysRevD.11.2558, PhysRevD.11.566}, and is addressed in various ways in beyond-the-SM theories.  
In models with extra spatial dimensions, the hierarchy problem can be solved by lowering the 
fundamental Planck scale to the TeV range, which implies that gravity is strongly coupled at the 
LHC energies. In this case, the production of microscopic black holes may be possible at the 
LHC~\cite{Dimopoulos:2001hw,Giddings:2001bu}. In some models~\cite{Meade:2007sz}, these black 
holes would decay nonthermally into pairs of particles. Such black holes are often called quantum 
black holes (QBHs).

The shape of the \mT distribution is interpreted using a binned-likelihood method for the SSM, QBH, and EFT descriptions.
We set upper exclusion limits on the cross section times branching fraction ($\sigma\mathcal{B}$) for the SSM description as a 
function of the mass of the new resonance 
($m_{\PWpr}$) and for the QBH model as a function of the mass scale of the QBH ($m_{\text{QBH}}$).
In the case of the EFT description, upper exclusion limits are placed on the Wilson 
coefficients.  
Additionally, a model independent limit setting method is considered and discussed.

The paper is structured as follows: 
the CMS detector is briefly described in Section~\ref{sec:CMSdetector}.
The models examined in this analysis are outlined in 
Section~\ref{sec:Signals}.
The data and simulated samples used are presented in Section~\ref{sec:Samples}, while 
Section~\ref{sec:Selection} describes 
the event reconstruction and the kinematic selection of the signal region. 
The specific determination of the background from misidentified \PGt leptons is described in Section~\ref{sec:background}.
Section~\ref{sec:systematics} 
reviews the most important sources of systematic uncertainties and their impact on the 
measurements.
The results and achieved sensitivity for all the models are detailed in 
Section~\ref{sec:results} 
and summarized in Section~\ref{sec:summary}.
The results are also documented in HEPData~\cite{HEPData}.

\section{The CMS detector}
\label{sec:CMSdetector}

The central feature of the CMS apparatus is a superconducting solenoid of 6\unit{m} internal diameter, providing a 
magnetic field of 3.8\unit{T}. 
Within the solenoid volume are a silicon pixel and silicon strip tracker, measuring the charged particle trajectories 
inside a pseudorapidity ($\eta$) acceptance of $\abs{\eta} < 2.5$. 
Also within the solenoid volume are a lead tungstate crystal electromagnetic calorimeter (ECAL), and a brass and 
scintillator hadron calorimeter (HCAL).
Each subdetector is composed of a barrel and two endcap sections. 
Forward calorimeters extend the pseudorapidity coverage provided by the barrel and endcap detectors. 
Muons are detected in gas-ionization detectors embedded in the steel flux-return yoke outside the solenoid.

Events of interest are selected using a two-tiered trigger system. The first level, 
composed of custom hardware processors,  uses information from the calorimeters and muon detectors to select events at a 
rate of around 100\unit{kHz} within a fixed latency of about 4\mus~\cite{Sirunyan:2020zal}. The second level, known as the 
high-level trigger, consists of a farm of processors running a version of the full event reconstruction software optimised 
for fast processing, and reduces the event rate to around 1\unit{kHz} before data storage~\cite{Khachatryan:2016bia}.

A more detailed description of the CMS detector, together with a definition of the coordinate system used and the relevant 
kinematic variables, can be found in Ref.~\cite{Chatrchyan:2008zzk}.

\section{Physics models}
\label{sec:Signals}

This section describes the different signal models used in this paper.
All of these models predict physics beyond the SM that would be detectable in a final state with a \PGt lepton 
and missing transverse momentum.

\subsection{The SSM heavy charged vector boson}
\label{ssec:wprime}

\begin{figure}[ht]
    \centering
    \includegraphics[width=0.49\textwidth]{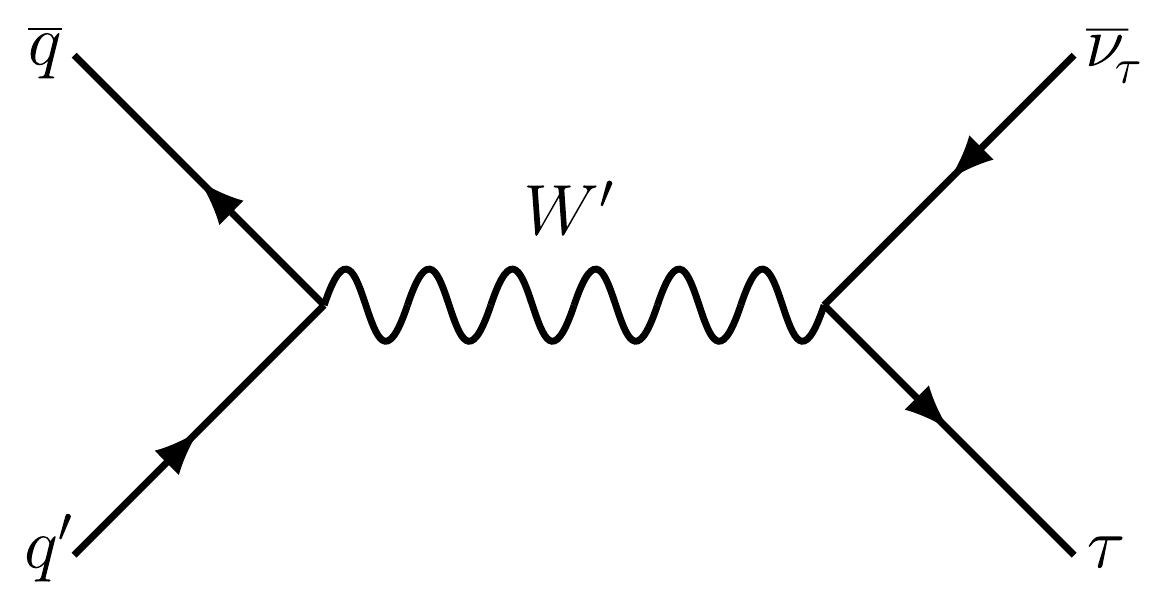}
    \caption{Leading order Feynman diagram for the production and decay of a new heavy charged vector 
boson, \PWpr, decaying to $\PGt\Pagngt$.}   
    \label{fig:Wprime_feynman}
\end{figure}

In the SSM, the \PWpr boson is a heavy analogue of the \PW boson. 
The experimental signature would be  a resonance with fermionic decay modes and branching fractions similar to 
those of the SM \PW boson.
For \PWpr boson masses larger than 180\GeV the additional $\PWpr\to\PQt\PQb$ decay 
mode becomes relevant. 
If this decay is considered, the branching fraction for the $\PWpr\to\PGt\PGn$ decay
in the SSM  is 8.5\%~\cite{Altarelli:1989ff}. 
Under these assumptions, the relative decay width of a 1\TeV \PWpr boson is $\Gamma/M \approx 3.3\%$. 
Decays to \PW and \PZ bosons are assumed to be suppressed in the SSM. 
No interference between the production of the new particle and the production of the SM \PW boson is assumed. 
This is justified by the fact that the subsequent hadronic decay of the \PGt lepton leads to two 
neutrinos being present in the final state, significantly lowering 
the effect an interference of \PWpr and \PW might have.
The leading order (LO) Feynman diagram for this specific process is given in Fig.~\ref{fig:Wprime_feynman}.

\subsection{Varying coupling strength}
\label{ssec:couplingmodel}

In the SSM, the \PWpr boson coupling strength, \WpCoupling is assumed to be identical to the SM \PW 
boson coupling, \Wcoupling.   
However, we also consider cases where $\WpCoupling \neq \Wcoupling$.
Because the \PWpr~boson width and cross section also depend on $\WpCoupling$, we can 
interpret the kinematic distributions with respect to \PWpr particles as described by various coupling 
values. 
The \PWpr boson coupling strength is given in terms of the SM weak coupling strength $ \Wcoupling = e/\swsq \approx 0.65$, 
where $e$ is the electron charge and $\theta_{\text{W}}$ is the weak mixing angle. 

\subsection{Nonuniversal gauge interaction model}
\label{ssec:NUGIMmodel}

Models with nonuniversal couplings predict an enhanced branching fraction for decays into the third generation of SM fermions.
These models could offer an explanation as to why the mass of the top quark is so 
large~\cite{Chiang:2009kb,Kim:2011qk,Edelhauser:2014yra}.   
The NUGIM exhibits an $SU(2)_{\ell} \times SU(2)_{\text{h}} \times 
U(1)$ symmetry.
The symmetry groups $SU(2)_{\ell}$ and $SU(2)_{\text{h}}$ govern the couplings of the gauge bosons to the light fermions 
of the first two generations and to the heavy fermions of the third generation, respectively.
Models of this type are often called G(221) models. 
The weak SM $SU(2)_{\text{L}}$ group is a low-energy limit of two gauge groups: $SU(2)_{\ell}$ and $SU(2)_{\text{h}}$. 

\begin{figure}[!htp]
    \centering
    \includegraphics[width=0.49\textwidth]{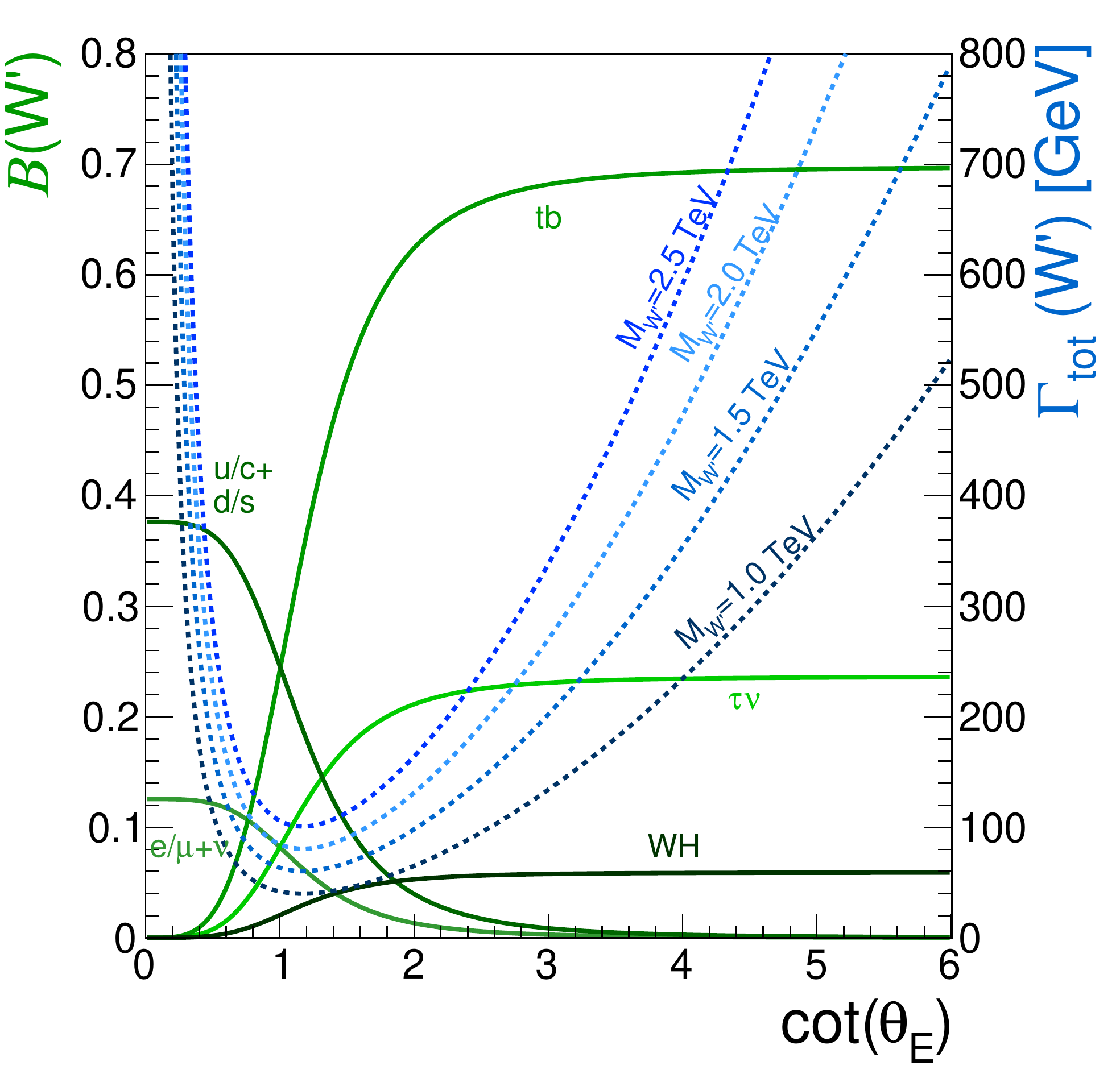}
    \caption{Branching fractions $\mathcal{B}(\PWpr)$ as function of the mixing angle $\cot\theta_{\text{E}}$ (solid lines), 
for \PWpr boson decays in the NUGIM G(221) framework, as calculated in 
Refs.~\cite{Chiang:2009kb,Kim:2011qk,Edelhauser:2014yra}. The total decay width
 $\Gamma_{\text{tot}}$ (dotted lines) can 
also be determined as a 
function of the mixing angle. When re-scaled to accommodate the $\PW\PH$ decay channel, the values associated 
with $\cot\theta_{\text{E}} = 1$ correspond to those in the SSM.        
    }
    \label{fig:NugimMixing}
\end{figure}

These groups combine via a mixing angle $\theta_{\text{E}}$, resulting in an SM-like $SU(2)_{\text{L}}$ group and an 
additional extended group $SU(2)_{\text{E}}$. 
This extended group gives rise to additional heavy gauge bosons such as the \PWpr boson. 
The mixing angle directly affects the cross section and decay width of such particles, as shown in 
Fig.~\ref{fig:NugimMixing}. 
As can be seen, for $\cot{\theta_{\text{E}}}=1$ the branching fractions are equal for all lepton flavors, 
which would correspond to an SSM-like feature. When re-scaled to accommodate the $\PW\PH$ decay channel, the values 
associated 
with $\cot\theta_{\text{E}} = 1$ correspond to those in the SSM.
For values larger than one, the decay to \PGt leptons becomes dominant. 
The mixing angle changes the production cross section and the branching fractions of the \PWpr boson. 
In the NUGIM, decays to \PW and \PZ bosons are negligible by construction, while decays to $\PW\PH$ bosons typically have a 
branching fraction ${<}6\%$. 

In the NUGIM G(221), neglecting the small contribution from the $\PW\PH$ decay, the ratio of the couplings, $g_{\PWpr}/g_{\PW}$, 
is related to the parameter $\cot\theta_{\text{E}}$
by the following equation~\cite{Edelhauser:2014yra}:
\begin{equation}
    \Gamma_{\PWpr}=\Gamma_{\PWpr}^{\text{SSM}}
    \frac{(4+\frac{1}{4})\cot^2\theta_\text{E}+8\tan^2\theta_\text{E}}{12+\frac{1}{4}}
    =\Gamma_{\PWpr}^{\text{SSM}}\left(\frac{g_{\PWpr}}{g_{\PW}}\right)^2.
\end{equation}
Because of this functional relationship, a reinterpretation of the limits on the coupling strength 
yields limits on the parameters of NUGIM G(221).   
Therefore, the SSM \PWpr signal samples mentioned in Section~\ref{sec:DataInput} can be used to probe this model.
The product of the cross section and branching fraction, $\sigma \mathcal{B}$, scales with $\Gamma_{\text{SSM}}/\Gamma_{\text{NUGIM}}$.

\subsection{Quantum black holes}
\label{ssec:quantumblackhole}

Theories that invoke extra spatial dimensions often permit a fundamental Planck scale in the TeV region. 
Such theories also provide the possibility of producing microscopic black holes at LHC 
energies~\cite{Dimopoulos:2001hw,Giddings:2001bu}. 
In models with QBHs~\cite{Meade:2007sz}, the regime at which gravity becomes strongly coupled to SM particles causes the 
theory to become non-perturbative.
This creates unavoidable limitations.
When working in these regimes, usage of the model and its predictions should be viewed as a dimensional analysis in which some 
extrapolations from the classical domain carry 
over to the QBHs.  
In contrast to semiclassical thermal black holes that can decay to high-multiplicity final states, QBHs are nonthermal 
objects, expected to decay predominantly to 
pairs of particles.  
Despite not having a complete theory of quantum gravity, it is still possible to gain insight into the signatures of QBHs 
at the LHC based on some fundamental 
principles and a few assumptions.    
The cross section for QBH production depends on the threshold mass, ${\text{m}}_{\text{th}}$, in $n$ additional spatial 
dimensions. 
The $n > 1$ cases correspond to the Arkani-Hamed--Dimopoulos--Dvali model~\cite{Arkani_Hamed_1998}.
In this search the number of extra dimensions is assumed to be n=4.
The ${\text{m}}_{\text{th}}$ distribution of the QBH signal is characterized by a sharp edge at the threshold of QBH production, followed by a 
monotonic decrease at larger masses.

\subsection{Leptoquarks}
\label{ssec:leptoquarks}

Leptoquarks are hypothetical new scalar or vector bosons carrying both baryon and lepton numbers and 
an electric charge in multiples of $(1/3)\,e$. 
The postulation of \TeV-scale LQs with dominant couplings to third-generation fermions is a popular 
theoretical explanation of the observed anomalies 
in
the ratio of the widths of semileptonic \PB meson decays 
\begin{equation}
  R_{\PD^{(*)}}=\frac{\Gamma(\PB\to \PD^{(*)}\PGt\PGn)}{\Gamma(\PB\to \PD^{(*)}\Pell\PGn)}
\label{equ:LQ-dstar}
\end{equation}
where $\Pell=\Pe$ or \PGm.

Under the assumption that their couplings to the SM 
fermions are renormalizable and gauge invariant, the possible quantum numbers of LQs are restricted to those given 
in the Buchm\"uller--R\"uckl--Wyler model~\cite{Buchmuller:1986zs}.       

\begin{figure}[htp]
  \centering
    \includegraphics[width=0.33\textwidth]{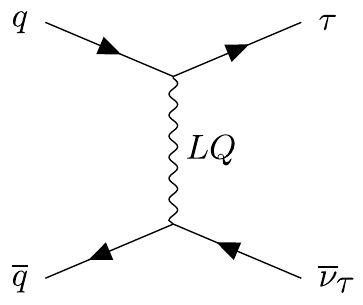}
   \caption{Leading order Feynman diagram of the process $\Pp{}\Pp\to\PGt\Pagngt$ 
mediated via a leptoquark in the $t$-channel.}   
  \label{fig:lqtch_generic_paper}
\end{figure}

At the LHC, the final state with one $\PGt$ lepton and \ptmiss can be produced by an LQ 
appearing as a $t$-channel propagator, leading to a nonresonant enhancement of events with large values 
of \mT. The LO Feynman diagram of this process is shown in 
Fig.~\ref{fig:lqtch_generic_paper}. 
The cross section of this process depends on the mass of the LQ ($m_{\text{LQ}}$) and its coupling 
$g_{\text{U}}$ to SM fermions. While the dependence of the cross section on $m_{\text{LQ}}$ is weaker for a LQ in the $t$-channel compared to single 
or pair LQ production, it scales with $g_{\text{U}}^4$ at LO.

In models addressing the tension in the flavor sector, LQs are often predicted to have flavor-nonuniversal 
couplings to left- (LH) and right-handed (RH) fermions, effectively modifying the LQ coupling to LH and RH SM 
fermions to be $g_{\text{U}}\beta_{\text{L}}^{ij}$ and $g_{\text{U}}\beta_{\text{R}}^{ij}$, respectively.
Here, $\beta_{\text{L}}^{ij}$ and $\beta_{\text{R}}^{ij}$ are complex $3{\times}3$ matrices with $i$ and $j$ 
denoting the down-type quarks and leptons, respectively. The normalization of $g_{\text{U}}$ is chosen such that 
$\beta_{\text{L}}^{ \PQb\PGt} = 1$.     

Three coupling benchmark scenarios are considered in this analysis, two of which have been proposed in 
Ref.~\cite{Cornella:2021sby}. The first two scenarios are the result of a fit of the parameters of 
a $U(1)$ vector LQ of 
charge $(-1/3)e$ to explain the flavor anomalies in 
$R_{\PD^{(*)}}$ 
(Eq.~(\ref{equ:LQ-dstar}))
and other 
related observables. The first benchmark scenario considers only couplings to LH SM fermions (\ie, 
$\beta_{\text{R}}^{ij} = 0$) and is referred to as ``best fit LH'' in the following.     
The second benchmark, referred to as ``best fit LH+RH'', additionally considers 
$\left\abs{\beta_{\text{R}}^{ \PQb\PGt}\right} = 1$ with all 
other $\beta_{\text{R}}^{ij} = 0$. In the third ``democratic'' benchmark, 
equal couplings only to LH fermions are assumed,
i.e. $\left\abs{\beta_{\text{L}}^{ij}\right} = 1$ and $\beta_{\text{R}}^{ij} = 0$ for all $i$ and $j$.

\subsection{Effective field theory description}
\label{ssec:EFTmodel}

As mentioned above, the ratio in Eq.~(\ref{equ:LQ-dstar}) appears to be enhanced with respect 
to the SM 
expectation 
by approximately 30\%.
If this enhancement originates from new physics, the corresponding mass scale of this physics is 
expected to be at, or near, the TeV scale~\cite{DiLuzio:2017chi}. 
The aim of this model is to search for any prescription that is able to address the 
$R_{\PD^{(*)}}$
anomalies with new 
heavy mediators. 
It uses a low-energy EFT of new physics in semi-leptonic $(\PQb\to\Pell\Pgn\PQc)$ 
transitions and translates this into a predictive theory at the TeV scale~\cite{Greljo:2018tzh}.    
This EFT model introduces three Wilson coefficients --- 
one for the LH vector coupling $\epsilon^{\text{cb}}_{\text{L}}$, one for tensor like coupling 
$\epsilon^{\text{cb}}_{\text{T}}$, and one for scalar-tensor like coupling $\epsilon^{\text{cb}}_{\text{S}_{\text{L}}}$.

\section{Data and simulated events}
\label{sec:Samples}

\subsection{Data}
\label{sec:DataInput}

In this search, we analyze a data set from pp collisions at a center-of-mass energy of 13 TeV 
that was recorded by the CMS experiment in 2016--2018 and  corresponds 
to an integrated luminosity of 138\fbinv. 
Events were selected with an isolated single \PGt trigger with $\abs{\eta} < 2.1$ and with \pt thresholds
of 120\GeV in 2016 and 180\GeV in 2017--2018. The \PGt trigger is 
complemented with a \ptmiss trigger with a 120\GeV threshold. The trigger efficiency for simulated signal events that pass the 
offline selection criteria is about 95\%. 
The efficiency of the \PGt trigger estimated using Monte Carlo simulations typically differs from that measured with data by 
approximately 10\% (depending on the particular \pt bin) across all data-taking years. 
The estimated \ptmiss trigger efficiency from simulation is only 1--4\% above or below data depending on the 
\ptmiss bin. 
We correct these differences in trigger efficiency between simulation and data using efficiency 
correction factors, which are henceforth referred to as scale factors.   
The scale factors range from  0.88--0.99 (0.87--1.08) for the 2016 (2017--2018) datasets.

\subsection{Event simulation}

Generated events are processed through a simulation of the detector geometry and response using
\GEANTfour~\cite{Agostinelli:2002hh}.
The same reconstruction software is applied to both data and simulated events.
Backgrounds are simulated using an expected distribution of the number of additional pp interactions within the same or nearby bunch crossings (pileup).
All simulated backgrounds are reweighted such that the distribution of the number of collisions per bunch crossing
matches the data, with an average of approximately 23 (32) interactions per bunch
crossing~\cite{CMS:2021xjt, CMS-PAS-LUM-17-004, CMS-PAS-LUM-18-002} in 2016 (2017--2018). Unless otherwise noted, simulated samples use the underlying event tune CUETP8M1~\cite{Skands:2014pea} in 2016 and the CP5 tune~\cite{CP5} in 2017--2018.

\subsubsection{Signal simulation}

The SSM \PWpr boson signal samples ($\PQq'\PAQq\to$\PWpr) are generated at LO using 
\PYTHIA8.2.05~\cite{Sjostrand:2014zea} for each data-taking
year, and employing the NNPDF2.3~\cite{Ball:2012cx} (NNPDF3.1~\cite{NNPDF:2017mvq}) parton distribution function
(PDF) set for the 2016 (2017--2018) configurations. 
These samples are generated for mass points ranging from 0.6 to 6.0 TeV. 
To interpret the results in terms of the coupling ratio of the SSM \PWpr boson and the SM \PW boson, signal 
samples are simulated at LO using \MGvATNLO version 2.6.5~\cite{madgraph5,Alwall:2014hca} for different coupling ratios 
$g_{\PWpr}/g_{\PW}$ 
for coupling ratio values of [0.001, 0.01, 0.1, 1.0, 2.0, 3.0, 5.0], for all mass points.   

Signal samples of LQ $t$-channel exchange at LO using 
\MGvATNLO are generated for LQ masses between 0.1 and 10\TeV 
neglecting interference with SM 
processes. 
Interference is expected to be negligible in the phase space sensitive to this 
signal~\cite{cmscollaboration2023search}.
The parton shower, hadronization, and underlying event are modelled with \PYTHIA8. 
The underlying event tune CP2~\cite{CP5} is used for 2017 and 2018.
The NNPDF2.3 and 
NNPDF3.1 PDF sets are used for 2016 and 2017--2018, respectively.      

The QBH signal samples are generated with the QBH 3.0 generator~\cite{Gingrich:2009hj}. 
We consider the $\PQq'\PAQq$ production of a singly positively charged QBH with its subsequent 
decay to a \PGt lepton and a neutrino. Signal samples with minimum threshold masses from 400 to 11000 GeV are employed.  
The CTEQ6L1 PDF set~\cite{Pumplin:2002vw}, which is the recommended PDF for QBHs, is used. 
The definition of the $Q$ scale for PDFs is taken to be the inverse of the gravitational radius. 

Signal samples for the EFT model are generated using \MGvATNLO at LO, while simulation of the 
hadronization and \PGt lepton decay is performed using \PYTHIA8.  
The signal samples are generated with \MGvATNLO using the values 0.3, 1.0, and 0.3 for the Wilson 
coefficients 
($\epsilon^{\text{cb}}_{\text{L}}$, $\epsilon^{\text{cb}}_{\text{T}}$, $\epsilon^{\text{cb}}_{\text{S}_{\text{L}}}$), 
respectively, which are obtained by 
fitting them to the current values of $R_{\PD^{(*)}}$~\cite{Alonso:2016gym}.   
The event content of these signal samples can be rescaled to cover a wider range of coupling values. 
The CP2 tune is used in conjunction with NNPDF3.1 PDF set for generation of signal samples.

\subsubsection{Simulation of the background}
\label{sec:SimBackground}

Background processes for this analysis include off-shell SM \PW production, processes coming from the decay of top 
quarks (\ttbar), the Drell--Yan (DY) process, and diboson ($\PZ\PZ$, $\PW\PW$, $\PW\PZ$) processes.
These samples are produced with \PYTHIA8 (for SM \PW), \MGvATNLO (for
DY, $\PZ\PZ$, and $\PW\PZ$), and
\POWHEG 2.0~\cite{Nason:2004rx,Frixione:2007vw,Alioli:2010xd} (for \ttbar and $\PW\PW$).
Backgrounds are simulated using the NNPDF3.0 (NNPDF3.1) NNLO \cite{Buckley:2014ana}  PDF set for 2016 
(2017--2018). The simulated SM samples were validated in the low-mass region ($\mT < 1\TeV$).  

The SM background expected in this search consists of two different types of events: those where the
\PGt candidate originates from a gluon- or quark-initiated jet, and others where the \PGt originates from a
genuine \PGt lepton decay.
The former is derived from data, while the latter is estimated using simulation.

The dominant irreducible background for this analysis comes from SM $\PW\to\PGt\PGn$ production.
This background is generated at LO precision in quantum chromodynamics (QCD) with \PYTHIA8 for 
high-mass 
off-shell production.
On-shell production is generated at LO precision in QCD with \MGvATNLO, and the simulated jets are matched to the matrix 
element 
and parton shower produced by \PYTHIA using MLM~\cite{madgraph5} merging.
The differential cross section is reweighted as a function of the invariant mass of the SM \PW boson,
incorporating next-to-LO (NLO) electroweak (EW) and next-to-NLO (NNLO) QCD corrections.
The EW and QCD corrections can be combined by two different mathematical approaches: additive and
multiplicative~\cite{Balossini:2009sa}.
These approaches differ by approximately 5\%.
The applied correction factor uses the recommended additive approach~\cite{Andersen:2014efa}, which is parametrized using
a fit of a fourth order polynomial function.
Differences between this approach and the multiplicative approach are taken as a systematic uncertainty.
The calculation uses the \FEWZ~3.1 and \textsc{mcsanc}~1.01~\cite{mcsanc_v1, Bondarenko:2013nu} generators for 
the QCD and EW corrections, respectively.

\section{Event selection and particle identification}
\label{sec:Selection}

The particle-flow algorithm~\cite{Sirunyan:2270046} aims to reconstruct and identify each individual 
particle in an event, with an optimized combination of information from the various subsystems of 
the CMS detector. 
We obtain the photon energy from the ECAL measurement. 
Electron energy is determined from a combination of the electron momentum at the primary 
interaction vertex as determined by the tracker, the energy of the corresponding ECAL cluster, and 
the energy sum from all bremsstrahlung photons spatially compatible with originating from the 
electron track.     
Muon energy is obtained from the curvature of the corresponding track. 
Charged-hadron energy is determined from a combination of the momentum of the hadron measured in the tracker and matching 
ECAL and HCAL energy deposits, corrected for the response function of the calorimeters to hadronic showers. 
Neutral-hadron energy is obtained from the energy deposits in the ECAL and HCAL corrected in a similar way to that of the 
charged hadrons. 

For each event, hadronic jets are clustered from these reconstructed particles using the infrared- and 
collinear-safe anti-\kt algorithm~\cite{Cacciari:2008gp}
in the FASTJET software package~\cite{Cacciari:2011ma} with a distance parameter of 0.4. 
The jet momentum is determined as the vectorial momentum sum of each particle contained within the 
jet. 
From simulation, we find the jet momentum to be, on average, within 5--10\% of the true momentum over 
the entire \pt spectrum and detector acceptance.   
Pileup can contribute additional tracks and calorimetric energy depositions to the jet momentum. 
To mitigate this effect, we discard charged particles identified to be originating from pileup vertices 
and  apply an offset to correct for remaining contributions. 
Jet energy corrections are derived from simulation to bring the measured response of jets to be 
similar to that of particle-level jets, on average. 
In situ measurements of the momentum balance in dijet, photon+jet, \PZ+jet, and multijet events are used to account for any 
residual differences in the jet energy scale between data and simulation~\cite{Khachatryan:2016kdb}. 
The jet energy resolution, obtained from a dijet balance technique, is found to be 15--20\% at 
30\GeV, 10\% at 100\GeV, and 5\% at 1 TeV~\cite{Khachatryan:2016kdb}. 
We apply additional selection criteria to each jet to remove jets potentially dominated by anomalous contributions from 
various subdetector components or reconstruction failures~\cite{Sirunyan:2019kia}.

The missing transverse momentum vector \ptvecmiss is computed as the negative vector \pt sum 
of all of the particle-flow candidates in an event, and its magnitude 
is denoted as \ptmiss.    
The \ptvecmiss is modified to account for corrections to the energy scale of the reconstructed jets 
in the event~\cite{Sirunyan:2019kia}.   
Anomalous high-\ptmiss events can occur due to a variety of reconstruction failures, detector malfunctions, or 
non-collision backgrounds. 
We reject such events by using event filters designed to identify more than 85\% of the spurious high-\ptmiss events, with a 
mistagging rate less than 0.1\%~\cite{Sirunyan:2019kia}. 

The \tauh candidates are reconstructed using the Hadron-plus-Strip (HPS)
algorithm~\cite{Khachatryan:2062435} starting from a jet.  
To reconstruct the \tauh decay mode and its four momentum,
we consider all particles in an $\eta$--$\phi$ cone of radius $\DR = 0.5$ around the jet axis. 
This candidate is subsequently compared with quark or gluon-initiated jets, electrons, and muons. 
We apply corrections for the small differences in the \tauh energy between simulation and data. 

To distinguish between \tauh candidates originating from genuine \PGt leptons and other objects, 
we use the \textsc{DeepTau} identification~\cite{CMS:2022prd} . 
This identification method consists of a convolutional deep neural network 
that uses machine learning techniques and image 
recognition methods in order to discriminate genuine \tauh candidates from light leptons and jets. 
It combines high-level variables, such as the \PGt lepton isolation, with low-level variables such as specific energy 
deposits in CMS subdetectors. 
This method has been shown to provide excellent discrimination against these jets, electrons, and muons while retaining a 
high efficiency for genuine \PGt leptons. 
The identification efficiency of genuine \tauh candidates ($\pt > 100$ GeV) provided by the 
chosen 
identification method in this analysis is approximately $75\%$, while the probability of a jet being 
misidentified as a \PGt lepton is less than 0.7\%.  
Differences between data and simulation arising from the application of the identification method are corrected for using scale factors.
Corrections due to differences in \PGt identification performance and energy scale measurements are less than 5\%.

We select \tauh candidates with a 
$\ensuremath{p_{\mathrm{T}}^{\PGt_{h}}}\xspace >$ 130 (190)\GeV for data recorded in 2016 
(2017--2018). 
The \tauh candidate is required to be in the range $\abs{\eta} < 2.1$. 
Additionally, we require \tauh candidates to pass anti-jet, anti-electron, and anti-muon 
discriminators of the \textsc{DeepTau} algorithm in order to reduce their corresponding misidentification 
probabilities~\cite{CMS:2022prd}. 
The \ptmiss associated with these events is required to be at least 200\GeV.

In order to reduce the number of misidentified light leptons and to reduce background, we reject events 
containing loosely identified electrons~\cite{CMS:2020uim} or muons~\cite{CMS:2018rym} with $\pt\approx > 20$~GeV within the detector 
acceptance. 
Events containing more than one \tauh candidate fulfilling the identification criteria are also rejected. 
The \tauh decay products and the neutrino are expected to be back-to-back to each 
other for all signal processes.    
The neutrino created during the \PGt lepton decay is emitted in a similar direction to that of the \PGt lepton in the 
laboratory frame
 and therefore only slightly reduces the overall magnitude of the expected \ptvecmiss. 
Thus, we require  $0.7 < p_{\text{T}}^{\tauh} /{\ptmiss} < 1.3$ and 
$\abs{\Delta \phi(\vec{p}_{\text{T}}^{\tauh}, \ptvecmiss) } > 2.4$.   
The variable used to determine the signal yields for this analysis is the transverse mass, $m_{\text{T}}$, as shown in Eq.~(\ref{equ:mt}).

\section{Jet background estimation}
\label{sec:background}

We use data to estimate the backgrounds due to jets misidentified as \tauh candidates. 
In this analysis, this background contribution originates mostly from QCD multijet and Drell--Yan events. 
The signal-free control region comprises events with one well reconstructed light lepton and a 
\tauh candidate. 
These events are selected using single light lepton triggers. 
The misidentification rate is defined as the number of \PGt lepton candidates passing the full selection 
criteria divided by the number of those candidates that do not fulfill the isolation requirement.
This rate is calculated as a function of the \PGt lepton candidate \pt and the number of charged hadrons within the 
\PGt lepton candidate isolation cone ($\DR = 0.5$).   
Events where a \PGt lepton is matched to a generator-level \PGt lepton in simulation are 
subtracted from data.
The ratio is estimated at approximately 5$\%$ for \PGt lepton decays involving three charged hadrons and up to 30$\%$ for 
\PGt leptons that decay into one charged hadron. 
The rate is then applied as an event-weight to signal-like events where a \PGt lepton candidate does not fulfill 
the isolation requirement.
This application region primarily contains events originating from misidentified jets, with a small contamination of 
genuine \tauh candidates.   
This contamination is subtracted from the application region using input from simulation. 

This method is validated by estimating the ratio of simulated events in which jets are misidentified as 
\tauh candidates, applying it to signal-like events where a \PGt lepton candidate does not fulfill 
the isolation requirement, and comparing the results to the background prediction from simulation in the signal region.   
In order to account for the differences found with this validation, and to cover uncertainties arising from 
differences in the misidentification rate measurements from the individual sideband regions, corresponding uncertainties are applied.

\section{Systematic uncertainties}
\label{sec:systematics}

There are various sources of uncertainties that can affect the normalization or the shape of the \mT distribution.  
These affect either the signal or the background estimates.
We consider three types of systematic uncertainties: normalization, shape-based, 
and statistical.  

The dominant uncertainty that affects the shape of the \mT distribution arises from the
PDF sets used in simulating the background processes.
We calculate this uncertainty with the PDF4LHC procedure~\cite{Butterworth:2015oua}, 
using replicas of the NNPDF3.0 (NNPDF3.1) PDF set for 2016 (2017--2018). 
This uncertainty grows with increasing \mT and becomes as large as 50\% in the highest bin ($> 1.8$ TeV) of the \mT 
distribution. 

Another source of uncertainty comes from corrections to the measured \PGt lepton 
energy scale (1.5--4.0\% for $p_{\text{T}}^{\tauh} > 100\GeV$, depending on the decay mode),
 \PGt identification efficiency (up to 5\% for 1\TeV \PGt leptons), and \PGt trigger scale factors ($\approx 10\%$). 
 Uncertainties coming from shifting the jet energy scale and resolution are propagated to the calculation 
of \ptvecmiss.

For all backgrounds estimated at NLO, factorization and renormalization scales are each multiplied 
by three factors (0.5, 1.0, 2.0), with the combination (1.0, 1.0) being the nominal. The largest
deviations in the 
resulting event yield predictions, both up and down from the nominal value, are taken as systematic uncertainty. 
Variations where one scale is multiplied by 0.5 and the other by 2 are considered unphysical and are not 
considered.

Events in simulation are reweighted based on pileup differences between measured data and simulation. The 
total inelastic cross section of 69.2\unit{mb}~\cite{CMS:2018mlc} is varied by $\pm 5\%$ and the resulting final \mT
distribution is recalculated to estimate the uncertainty.

Uncertainties that affect the normalization of the background modeling arise from the measurement of the integrated 
luminosity, as well as from uncertainties in the cross sections of specific backgrounds. 
For the integrated luminosity, we apply an uncertainty of 1.2\%/2.3\%/2.5\% 
for the year 2016/2017/2018, respectively, which correspond to the combined uncertainty of 1.6\% for the full data-taking 
period. 
The cross section uncertainties are evaluated as 4\% for the diboson background and 2$\%$ for the Drell--Yan background. 
The background from misidentified \PGt lepton candidates is estimated with a normalization uncertainty of 100\%. 
Additionally, we estimate the systematic uncertainty in the calculated ratio arising from the
statistical uncertainties in the control and application regions.  
For the 
dominant \PW boson background, we apply the \PW boson mass-dependent uncertainties as explained in 
Section~\ref{sec:SimBackground}. 
Uncertainties in the jet background estimation include uncertainties from the misidentification ratio measurement, as well 
as the statistical uncertainties from both control and application regions of the method. 
Statistical uncertainties arise from data and from the limited event count in simulated background 
samples, and 
are taken into account~\cite{BARLOW1993219}. 
These uncertainties are treated as uncorrelated between the bins of the final \mT 
distribution. 
The combination of all uncertainties shows an increasing trend with increasing \mT.
Because the number of total expected 
background events becomes small, this does not affect the sensitivity by a significant amount for 
high-mass objects. 

All systematic uncertainties are treated as fully correlated across years, except for those where the uncertainty arises 
from statistically
independent sources. The uncertainties of the data-driven 
background are treated as uncorrelated across years. Correction uncertainties related to \tauh and \ptmiss and the 
luminosity uncertainties are treated as partially 
correlated since they include some components that are considered correlated between years and some others that are not.

\section{Results}
\label{sec:results}

\begin{figure}[htp]
\centering
\includegraphics[width=0.6\textwidth]{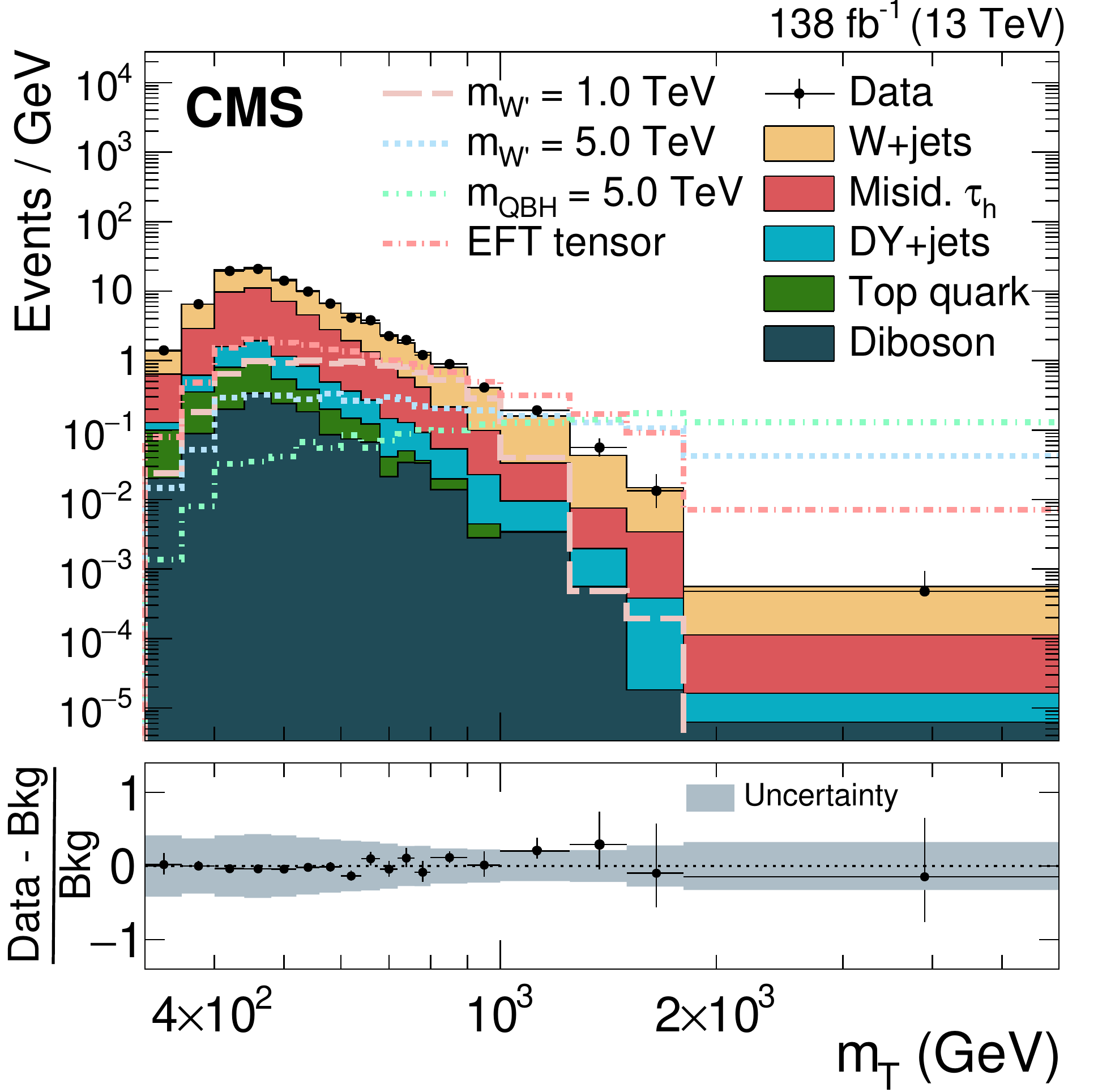} 
\caption{The transverse mass distribution of the $\tauh+\ptmiss$ system observed in 
the 2016--2018 data (black dots with statistical uncertainty) as well as the expectation from SM processes 
(stacked histograms). The horizontal lines of the data points reflect the varying bin sizes. Different signal hypotheses normalized to 10\fbinv are illustrated as dashed lines for exemplary
SSM \PWpr boson, QBH, and EFT signal hypotheses.
The ratios of the background-subtracted data yield to the expected background yield are presented in the lower panel. 
The combined statistical and systematic uncertainties in the background are represented by the grey shaded band in 
the ratio panel.}  
\label{fig:MT_final}
\end{figure}

Figure~\ref{fig:MT_final} displays the distribution for hadronically decaying \PGt lepton events with missing transverse 
momentum, for the combined 2016--2018 data set. 
To illustrate signal distributions for the models described in this analysis, simulated SSM \PWpr boson samples, as well 
as simulated QBH and EFT samples are included in the figure. 
The \PWpr boson masses, $m_{\mathrm{\PWpr}}$, of 1.0 and 5.0~\TeV are used for the SSM samples, 
a QBH mass, 
$\mathrm{m_{QBH}}$, of 5.0~\TeV is used for the QBH sample, and a tensor-type EFT sample is used to show the expected 
effect of the Wilson coefficients on the \mT distribution.   
The main background contribution comes from off-shell \PW boson production, followed 
by events with misidentified \PGt leptons. Backgrounds from misidentified \PGt leptons are estimated 
using the 
data-driven technique described in Sec.~\ref{sec:background}. Other backgrounds are estimated using simulation.    
Example signal distributions are normalized to a cross section of 10\unit{fb}. 
The bin width increases with transverse mass. 

The data agree with the expected background yields within statistical and systematic uncertainties. No significant 
deviation from the SM prediction is observed. 
We calculate limits at 95\% confidence level (\CL)  from the \mT distribution using Bayesian methods known to have good 
frequentist coverage properties~\cite{ParticleDataGroup:2022pth}.
A software framework designed specifically to combine Higgs boson searches~\cite{higgsCombine}, based on the 
\textsc{RooStats} package~\cite{RooStats}, is used to calculate the limits.
For these limit calculations, we use a Bayesian method with a uniform positive prior probability density for the signal 
cross section. 
The nuisance parameters connected to shape-based uncertainties are modeled using a template-morphing technique, while 
those affecting the normalization of the distribution are included using a log-normal prior.
To search for resonances in the \mT spectrum, a binned maximum likelihood fit is performed.

We calculate upper exclusion limits on $\sigma \mathcal{B}$
for each model described in Section~\ref{sec:Signals}.  
We further translate these into lower exclusion limits on $m_{\PWpr}$ and $m_{\text{QBH}}$ for the SSM \PWpr and QBH 
models, respectively.
However, for the studies focused on coupling strengths, these are left as upper exclusion limits, unless otherwise 
specified.

\subsection{Model-independent cross section limit}

\begin{figure}[!htp]
	\centering 
	\includegraphics[width=0.6\textwidth]{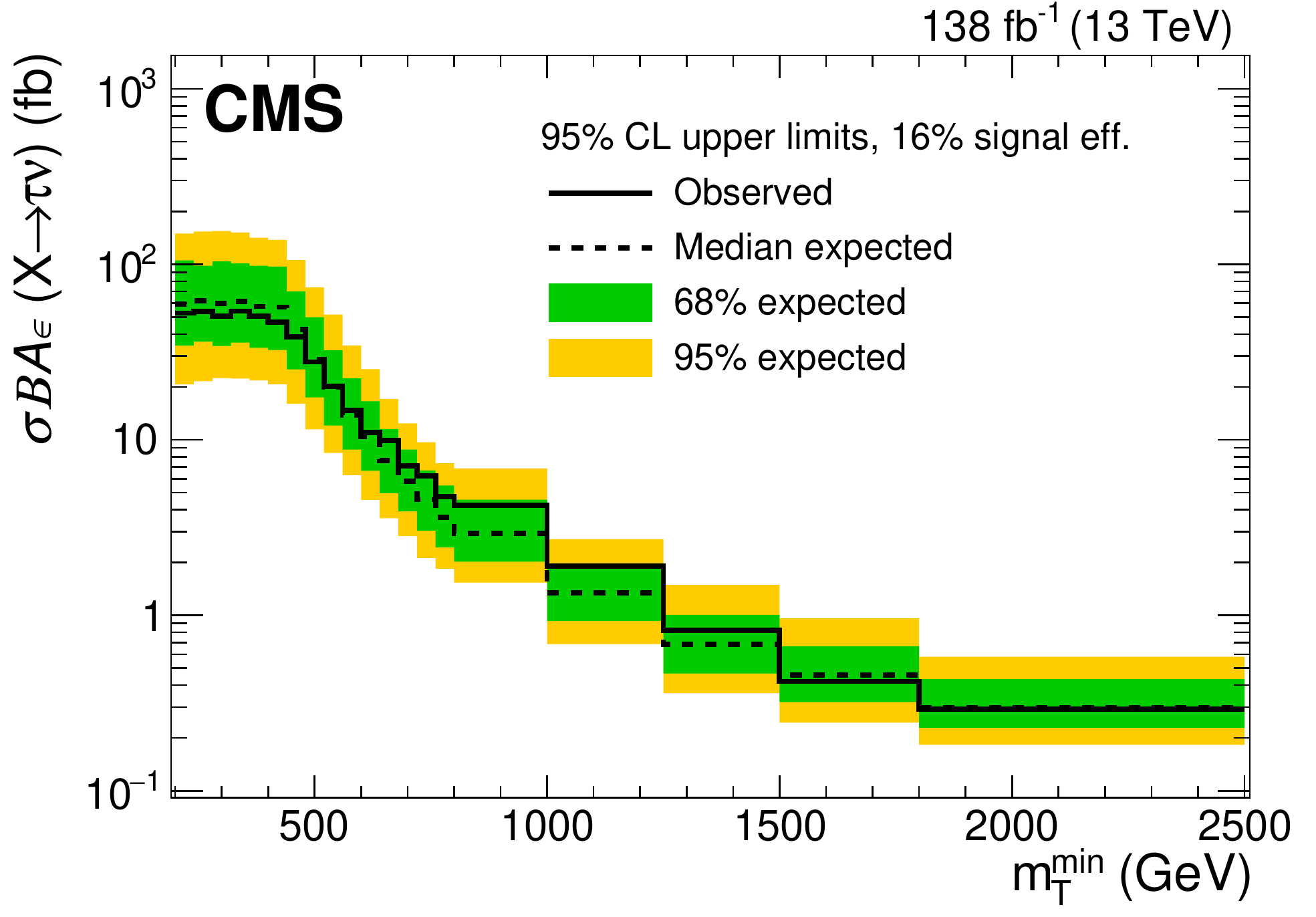}
	\caption{Bayesian 95\% \CL model-independent upper limit on the product of signal cross section and branching 
fraction for the \PGt+\PGn decay for a back-to-back \PGt lepton plus \ptmiss 
topology.   
To calculate this limit, all events for signal, background, and data are summed starting from a minimum \mT threshold and 
then divided by the total number of events. 
No assumption on the signal shape is included in this limit, however a signal selection efficiency of 16\% is assumed. 
The expected (dashed line) and observed (solid line) limits are shown as well as the 68\% and 95\% \CL 
uncertainty bands (green and yellow, respectively). The uncertainty bands correspond to the sum in quadrature of PDF and 
scale variations.}
	\label{fig:MILimit}
\end{figure}

New physics might produce signal shapes that are different from those expected for any of the above-mentioned signal models.   
Therefore, a model-independent (MI) cross section limit on $(\sigma \mathcal{B} A \epsilon)_{\text{MI}}$ is 
provided in Fig.~\ref{fig:MILimit}. This limit is obtained by calculating a single bin limit, where we integrate from 
a lower \mT threshold to infinity. This procedure results in an exclusion limit with a step-like
behavior that originates from the binning of the \mT distribution.
For the estimation of this exclusion limit, a constant product of acceptance times efficiency of $16\%$ is assumed. This value 
includes the branching fraction of hadronic \PGt lepton decays and experimental efficiencies for the signal. An exclusion limit 
on a specific model calculated using the presented MI limit can be obtained by applying the model-dependent part of $A \epsilon$. 

To reflect the effect of the threshold, $\mT^{\text{min}}$ on a new physics signal, we determine a factor $f_{\mT}$ 
by counting the number of events with $\mT > \mT^{\text{min}}$ and dividing this result by the number of generated events.  
The reconstruction efficiency is nearly constant over the entire \mT range scanned in this analysis;
therefore $f_{\mT}$ can be evaluated at the generator level. 

A limit on the product of the production cross section and branching fraction $(\sigma 
\mathcal{B} A 
\epsilon)_{\text{excl}}$ together with the acceptance $A$ and the efficiency $\epsilon$
can be obtained as a function of $\mT^{\text{min}}$ by dividing the excluded cross section of the MI limit 
$(\sigma 
\mathcal{B} A \epsilon)_{\text{MI}}$ given in Fig.~\ref{fig:MILimit} by the calculated fraction 
$f_{\mT}(\mT^{\text{min}})$:
\begin{equation}
(\sigma \mathcal{B} A \epsilon)_{\text{excl}} = \frac{(\sigma \mathcal{B} A 
\epsilon)_{\text{MI}}(\mT^{\text{min}})}{f_{\mT}(\mT^{\text{min}})}
\end{equation}

Here, $\mathcal{B}$ is the branching fraction of the new particle decaying to \PGt+\PGn. A new 
physics model can be 
interpreted using the MI limit by scanning $\mT^{\text{min}}$ for the minimum value 
of $(\sigma \mathcal{B} A \epsilon)_{\text{excl}}$. When accounting for the model-dependent part of $A \epsilon$, $(\sigma \mathcal{B})_{\text{excl}}$ can be extracted.
Models with a theoretical cross section $(\sigma \mathcal{B})_{\text{theo}} > (\sigma \mathcal{B})_{\text{excl}}$ can be 
excluded. 
The procedure described here can be applied to all models involving the two-body decay of a massive system to a \PGt 
lepton and \ptmiss, which exhibit back-to-back kinematics similar to those of a generic \PWpr boson. 
If the kinematic properties are different, the fraction of events  $f_{\mT}(\mT^{\text{min}})$ must be determined for the 
particular model considered.

\subsection{Mass limit for a heavy vector boson}

\begin{figure}[!htp]
	\centering 
		\includegraphics[width=0.6\textwidth]{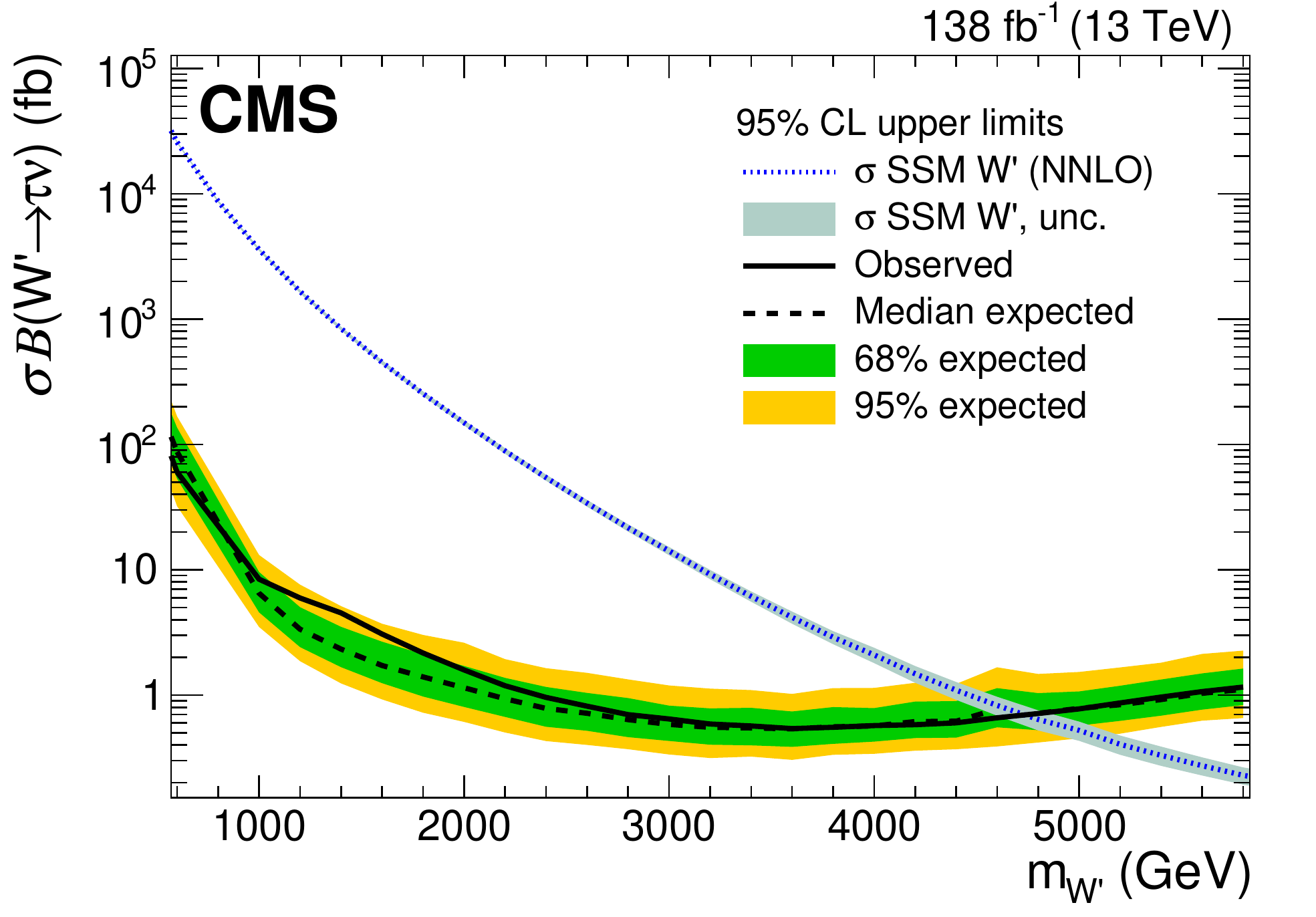}
	\caption{Bayesian upper exclusion limits at 95\% \CL on the product of the cross section and branching fraction of 
a \PWpr boson decaying to a \PGt lepton and a neutrino in the SSM model. 
For this model, \PWpr boson masses of up to 4.8\TeV can be excluded. 
The limit is given by the intersection of the observed (solid) limit and the theoretical cross 
section (blue dotted curve). 
The 68 and 95\% quantiles of the limits are represented by the green and yellow bands, respectively. 
The $\sigma \mathcal{B}$ for an SSM \PWpr boson, along with its associated uncertainty, calculated at NNLO precision in 
QCD is shown. }
	\label{fig:SSMLimit}
\end{figure}

When calculating the limit in the SSM, the parameter of interest is the product of the signal 
production cross section and the 
branching fraction $\sigma \mathcal{B}(\PWpr\to\PGt\PGn)$.  
Figure~\ref{fig:SSMLimit} shows the exclusion limit on this product as a function of the SSM $m_{\PWpr}$. 
The SSM \PWpr boson production is excluded for masses between 0.6\TeV and 4.8\TeV by comparison with the 
theoretical NNLO cross section. 
The observed limit is in agreement with the expected mass limit of 4.8\TeV. 
Previous CMS searches in this channel already exclude the lower SSM mass hypotheses~\cite{Khachatryan:2015pua}.
The lower bound is mainly determined by the trigger thresholds and the upper bound by the available data. 
In the high mass region, off-shell production of \PWpr bosons becomes dominant, shifting the signal \mT 
distribution towards lower \mT values.     

\subsection{The NUGIM and coupling sensitivity}

\begin{figure}[!htp]
	\centering 
		\includegraphics[width=0.49\textwidth]{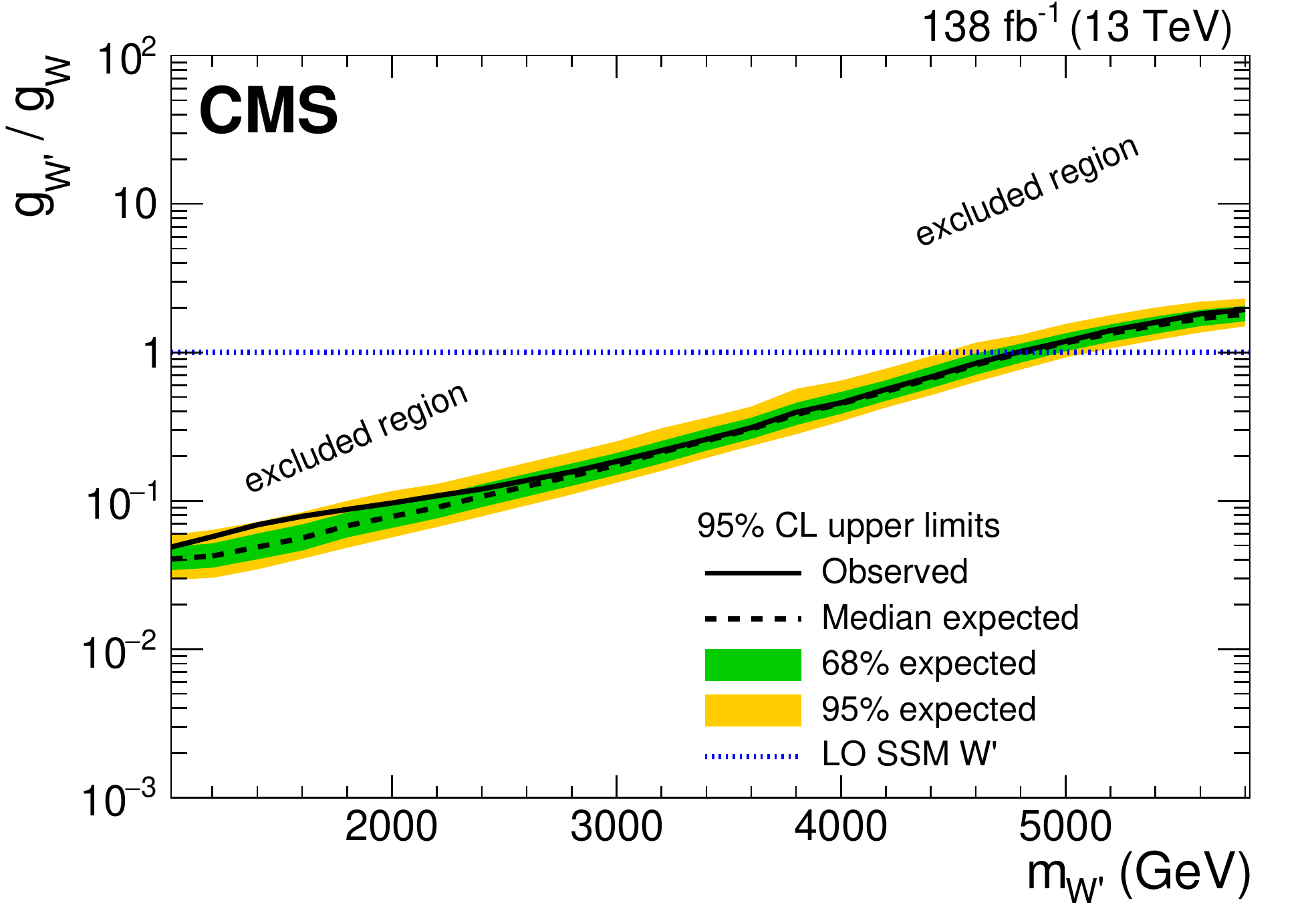}
		\includegraphics[width=0.49\textwidth]{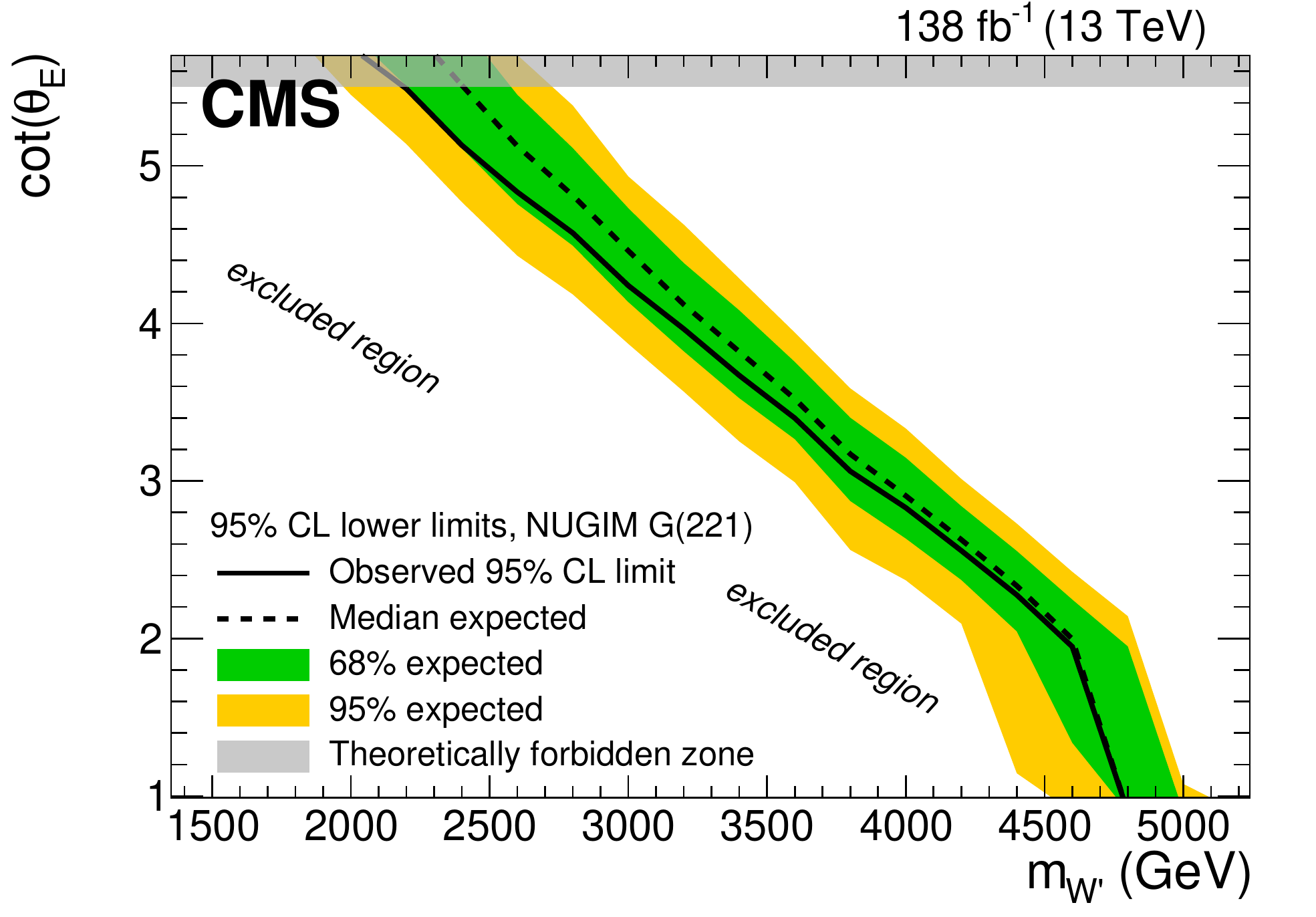}
	\caption{ 
Bayesian upper exclusion limits on the ratio $g_{\PWpr}/g_{\PW}$ for an SSM-like \PWpr boson are shown on the left. 
The 
unity coupling ratio (blue dotted curve) corresponds to the SSM common benchmark.    
The lower exclusion limits on the NUGIM G(221) mixing angle cot$\theta_{\text{E}}$
are shown on the right
as a function of the \PWpr boson mass. The theoretically excluded region is shaded in grey.
The 68 and 95\% quantiles of the limits are represented by the green and yellow bands, respectively.
}
	\label{fig:coupling_limits}
\end{figure}

The limits for the coupling-dependent \PWpr boson model and the NUGIM G(221) model are calculated in a two-dimensional 
plane that depends on the mass of the new mediator particle and its width. 
The \PWpr boson width is affected by the coupling. 
We extract the limits shown in Fig.~\ref{fig:coupling_limits} by finding the intersection of the 
LO cross section theory plane of the corresponding model with the experimental result. 
The upper limits on the coupling strength in the case of SSM-like \PWpr bosons, and on the mixing parameter 
$\cot\theta_{\text{E}}$ in the case of the NUGIM G(221) model are provided as functions of $m_{\PWpr}$.
In the latter case, for $\cot(\theta_{\text{E}}) > 5.5$, the width becomes so large that the model is no longer valid. 
We exclude coupling values above $g_{\PWpr}/g_{\PW} = 0.05$ for masses of 1\TeV. 
For a mass hypothesis of 6\TeV, coupling values above 2 are excluded. 
For NUGIM G(221) models, we can exclude all theoretically allowed values of $\cot{\theta_{\text{E}}}$ for masses below 2.2 
TeV. 
For values of  $\cot{\theta_{\text{E}}} = 1$, masses of up to 4.8\TeV are excluded.
 
\subsection{The QBH limits}
\begin{figure}[!htpb]
	\centering 
		\includegraphics[width=0.6\textwidth]{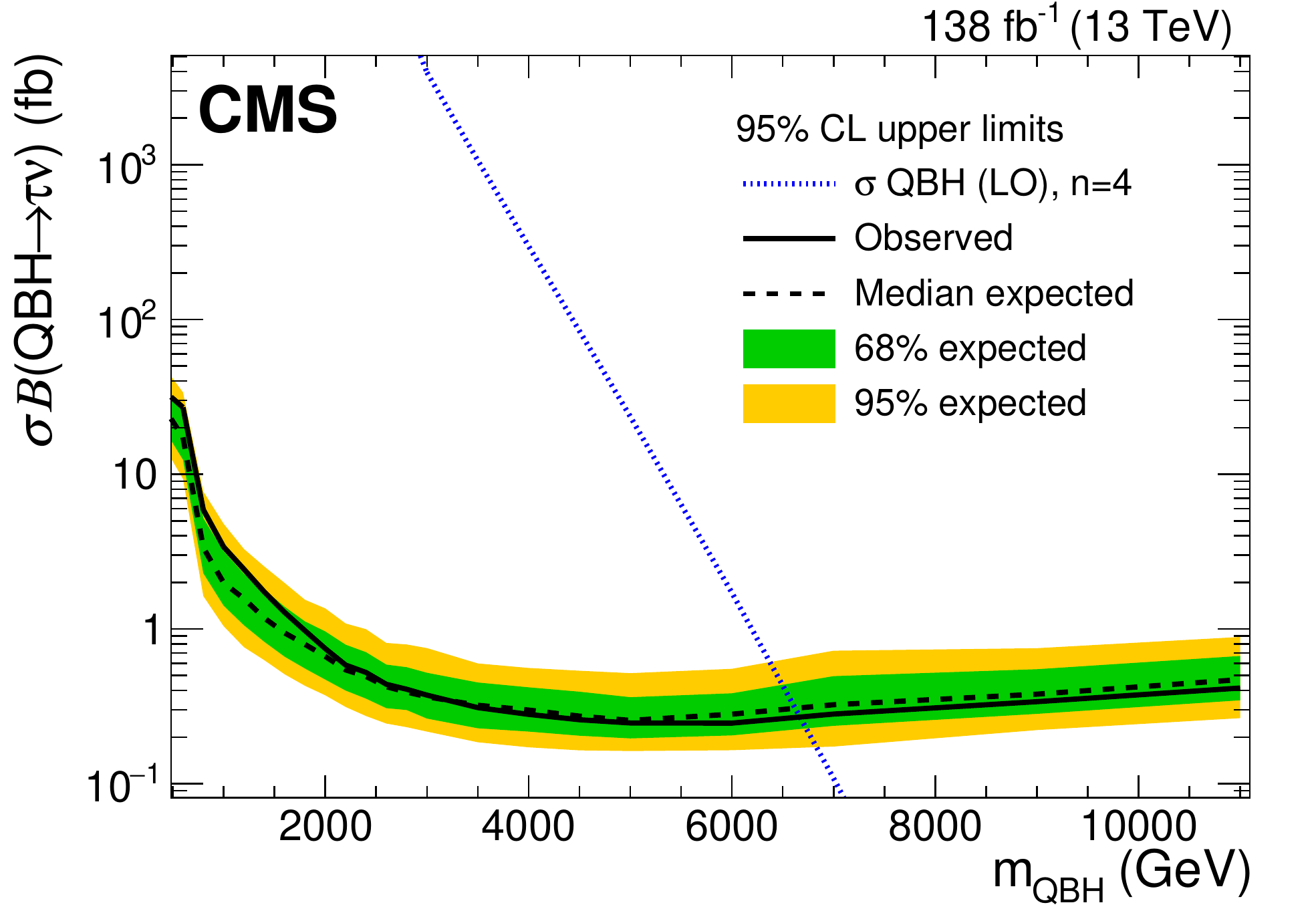}
	\caption{Bayesian upper exclusion limits at 95\% \CL on the product of the production cross section 
and branching fraction of a QBH in an associated \PGt lepton and neutrino final state. 
The 68 and 95\% quantiles of the limits are represented by the green and yellow bands, respectively.
Masses of up to 6.6\TeV are excluded at 95\% \CL. The limit is given by the intersection of the observed (solid) limit and the 
theoretical cross section (blue dotted curve). }
	\label{fig:QBHLimit}
\end{figure}

The QBH model has the same \mT sensitive observable as the SSM model.  However, it differs with respect to 
its production mechanism. 
Its signal shape in the generator mass distribution is characterized by a threshold mass with a rising edge, followed by a 
monotonic decrease towards higher masses. 
This shape is less sharp in the \mT distribution and, in parts, extends towards lower masses because of the \PGt lepton 
decay. 
We exclude this model for threshold masses of up to 6.6\TeV, as shown in Fig.~\ref{fig:QBHLimit}.

\subsection{Leptoquark limits}

\begin{figure}[t]
  \centering
    \includegraphics[width=0.6\textwidth]{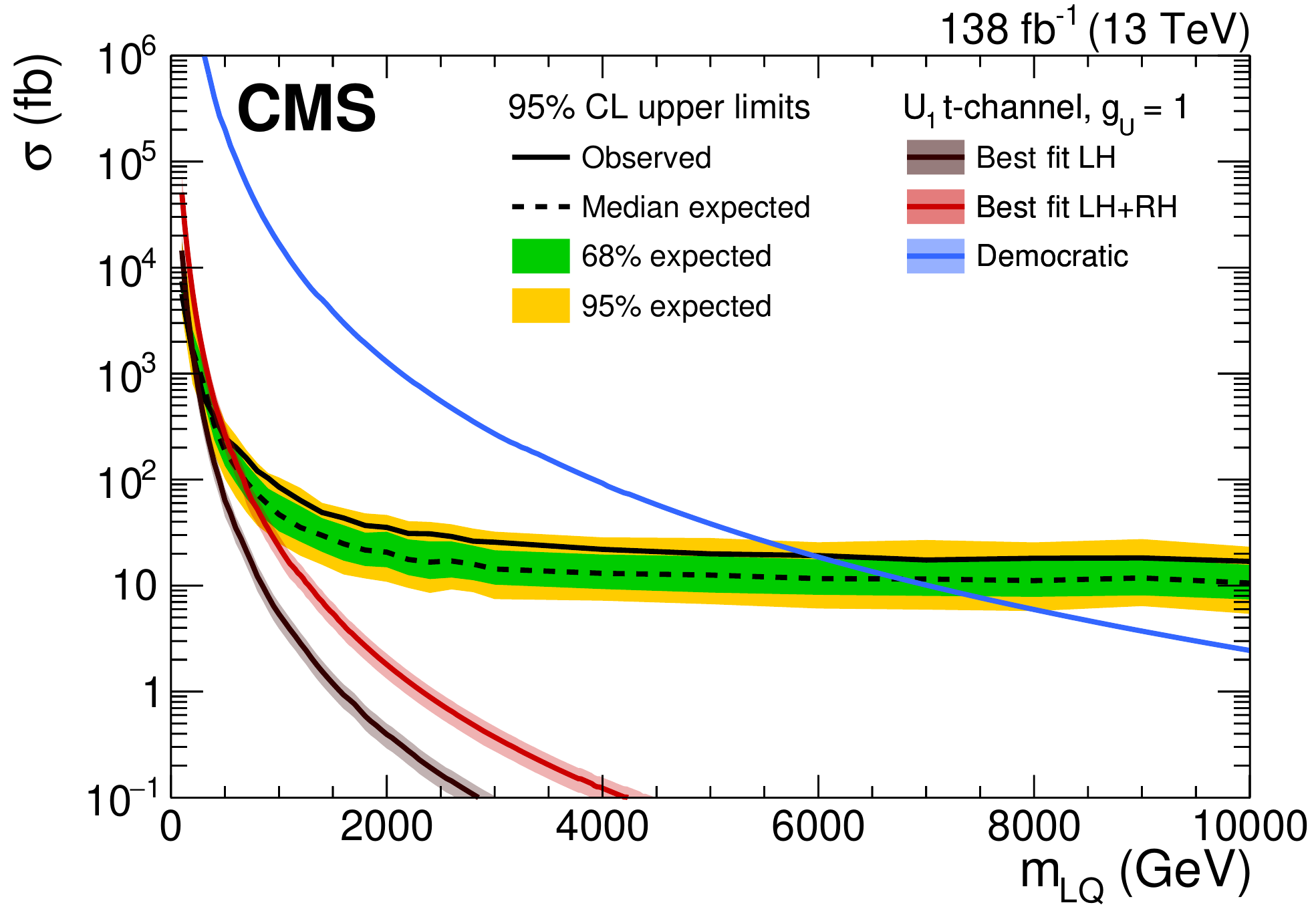}  
  \caption{Bayesian upper limits at 95\% \CL on the cross section of the process 
$\Pp{}\Pp\to\PGt\PGn$ mediated 
by LQ exchange in the $t$-channel.
The inner (green) band and the outer 
(yellow) band indicate the regions containing 68 and 95\%, respectively, of the distribution of limits expected 
under the background-only hypothesis. The predicted LQ cross section at LO in the three coupling benchmark 
scenarios is depicted in different colors for $g_{\mathrm{U}}=1$.
The uncertainty bands correspond to the sum in quadrature of PDF and scale variations.
The first benchmark scenario considers only couplings to LH SM fermions (i.e.
$\beta_{\text{R}}^{ij} = 0$) and is referred to as ``best fit LH''.
The second benchmark, referred to as ``best fit LH+RH'', considers $\left\abs{\beta_{\text{R}}^{ \PQb\PGt}\right} = 1$ and all
other $\beta_{\text{R}}^{ij} = 0$. In the third ``democratic'' benchmark,
equal couplings only to LH fermions are assumed,
\ie, $\left\abs{\beta_{\rm{L}}^{ij}\right} = 1$ and $\beta_{\rm{R}}^{ij} = 0$ for all $i$ and $j$.
}
  \label{fig:limits1d_paper}
\end{figure}

We also place upper limits  on 
the cross section of the process $\Pp{}\Pp\to\PGt\PGn$ mediated by LQ exchange in the $t$-channel. 
These limits are 
presented in Fig.~\ref{fig:limits1d_paper}. Masses up to 205/ 515 / 5900 GeV are excluded for the best-fit LH / best-fit 
LH+RH / democratic scenarios.

\begin{figure}[th]
  \centering
  \includegraphics[width=0.49\textwidth]{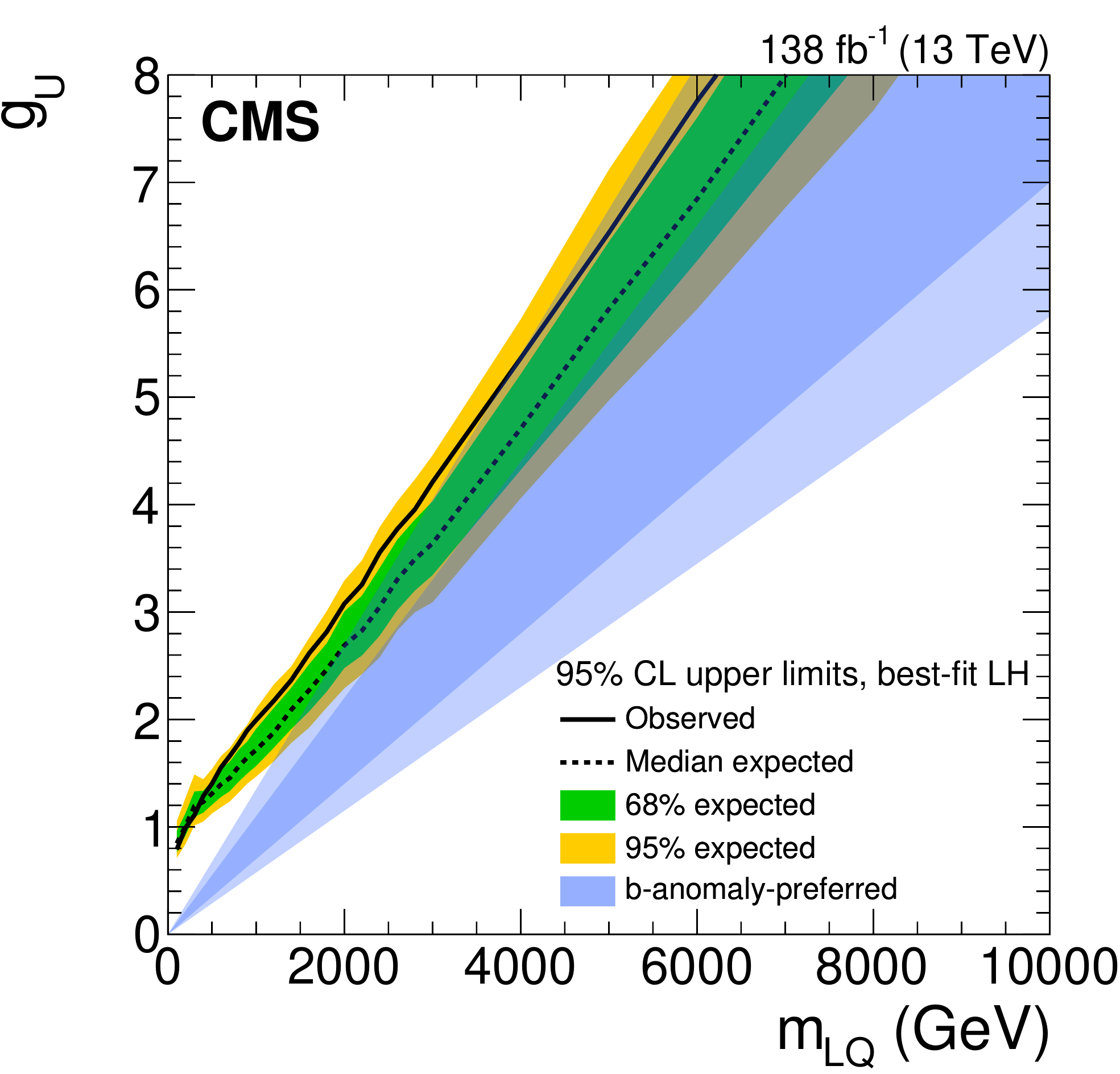}
  \includegraphics[width=0.49\textwidth]{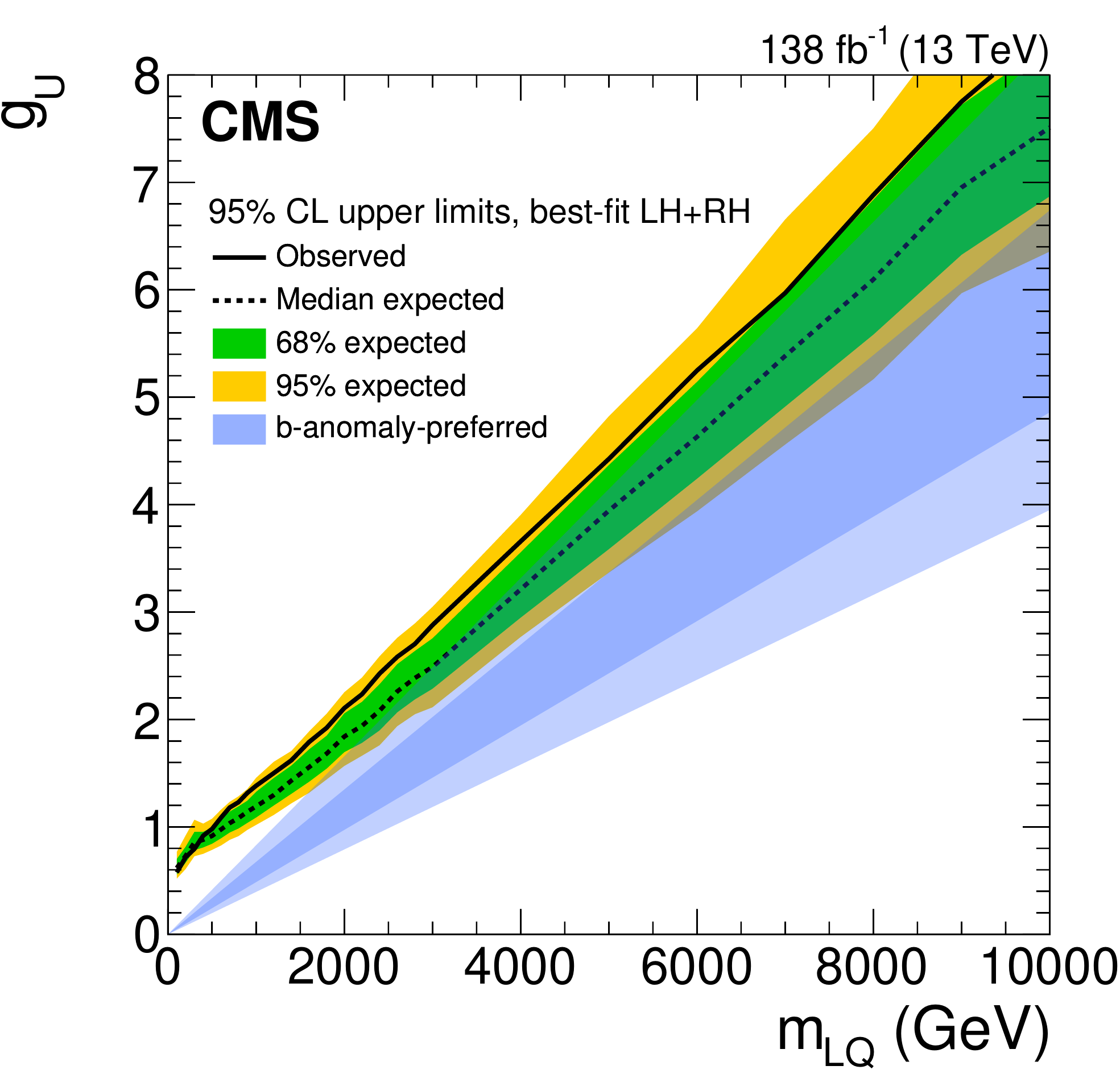}\\
  \includegraphics[width=0.49\textwidth]{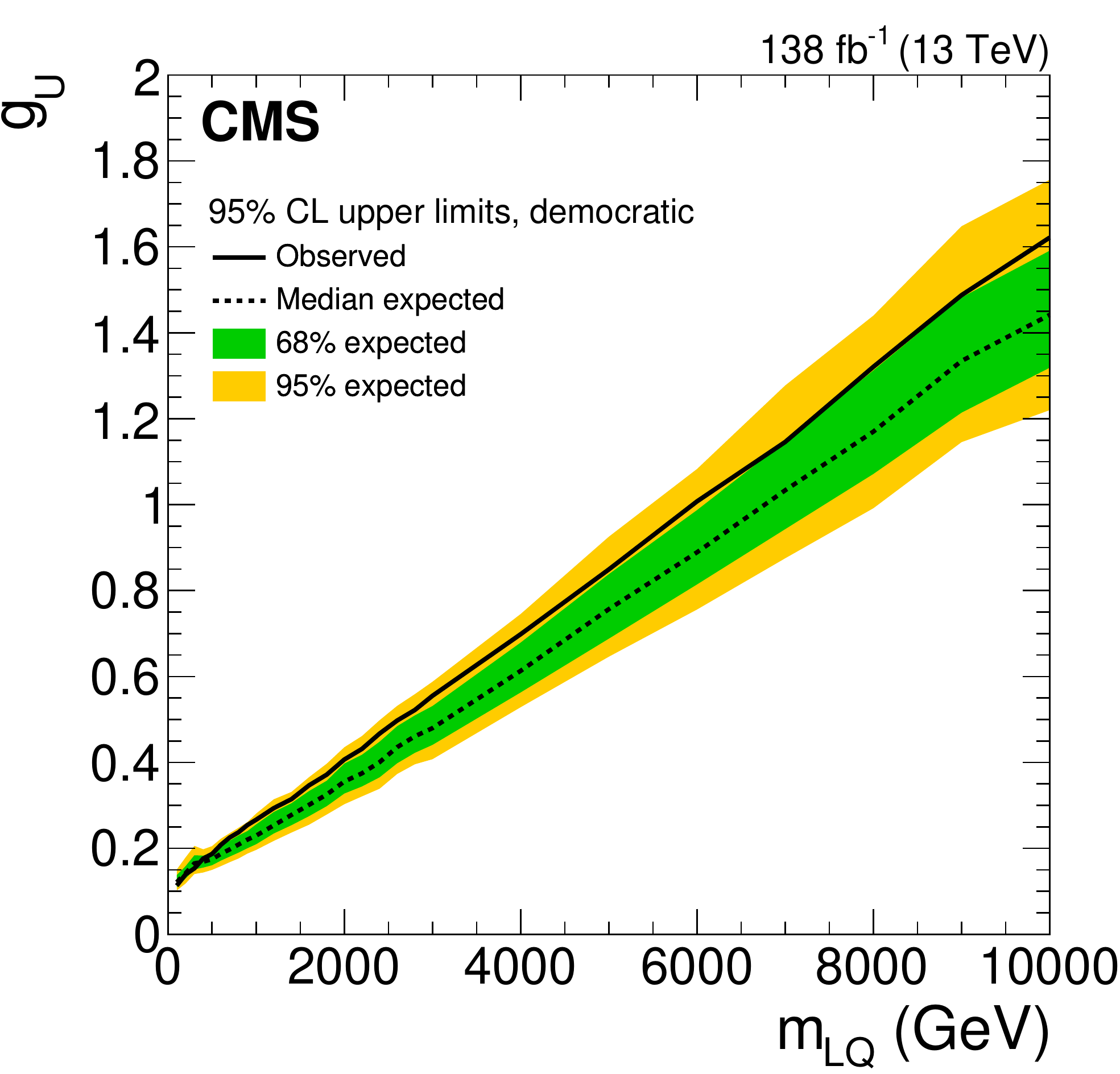}
  \caption{Expected and observed upper limits of the LQ coupling $g_{\mathrm{U}}$ as a function of the mass in the LH 
(left), LH+RH (right), and democratic (lower) scenarios. The blue band shows the 68 and 95\% regions of 
$g_{\mathrm{U}}$ preferred by the fit to the \PQb anomalies data~\cite{Cornella:2021sby}. 
}
  \label{fig:limits2d_paper}
\end{figure}

Figure~\ref{fig:limits2d_paper} shows lower limits on the LQ mass for varying LQ couplings $g_{\text{U}}$ to SM fermions.
The limits are compared to the values of $g_{\mathrm{U}}$ that are obtained in 
the fit~\cite{Cornella:2021sby} to the flavor anomaly data in the two best fit scenarios for only LH couplings 
(left) and LH+RH couplings (right). 

In both best-fit scenarios, we exclude a significant part of the LQ parameter space. LQ 
masses up to 10\TeV and couplings $g_{\text{U}}$ of up to 8 are probed. In comparison to the 
values of 
$g_{\text{U}}$ preferred to explain the flavor anomalies, this analysis has direct sensitivity to LQs, as 
predicted by the model of Ref.~\cite{Cornella:2021sby}. 
The exclusion limits at 95\% \CL just reach the preferred phase space at large values of $m_{\text{LQ}}$ and $g_{\text{U}}$.
In addition to exclusion limits 
probing the couplings as predicted in the best fit scenarios, limits that are not motivated by the \Pb anomalies are 
presented as well. 
They are obtained using the benchmark of democratic LH couplings. The lower limits on $m_{\text{LQ}}$ are 
shown for varying values of $g_{\text{U}}$ in Fig.~\ref{fig:limits2d_paper} (lower) for this scenario.

\subsection{The EFT limits}

\begin{figure}[!htp]
	\centering 
	\includegraphics[width=0.6\textwidth]{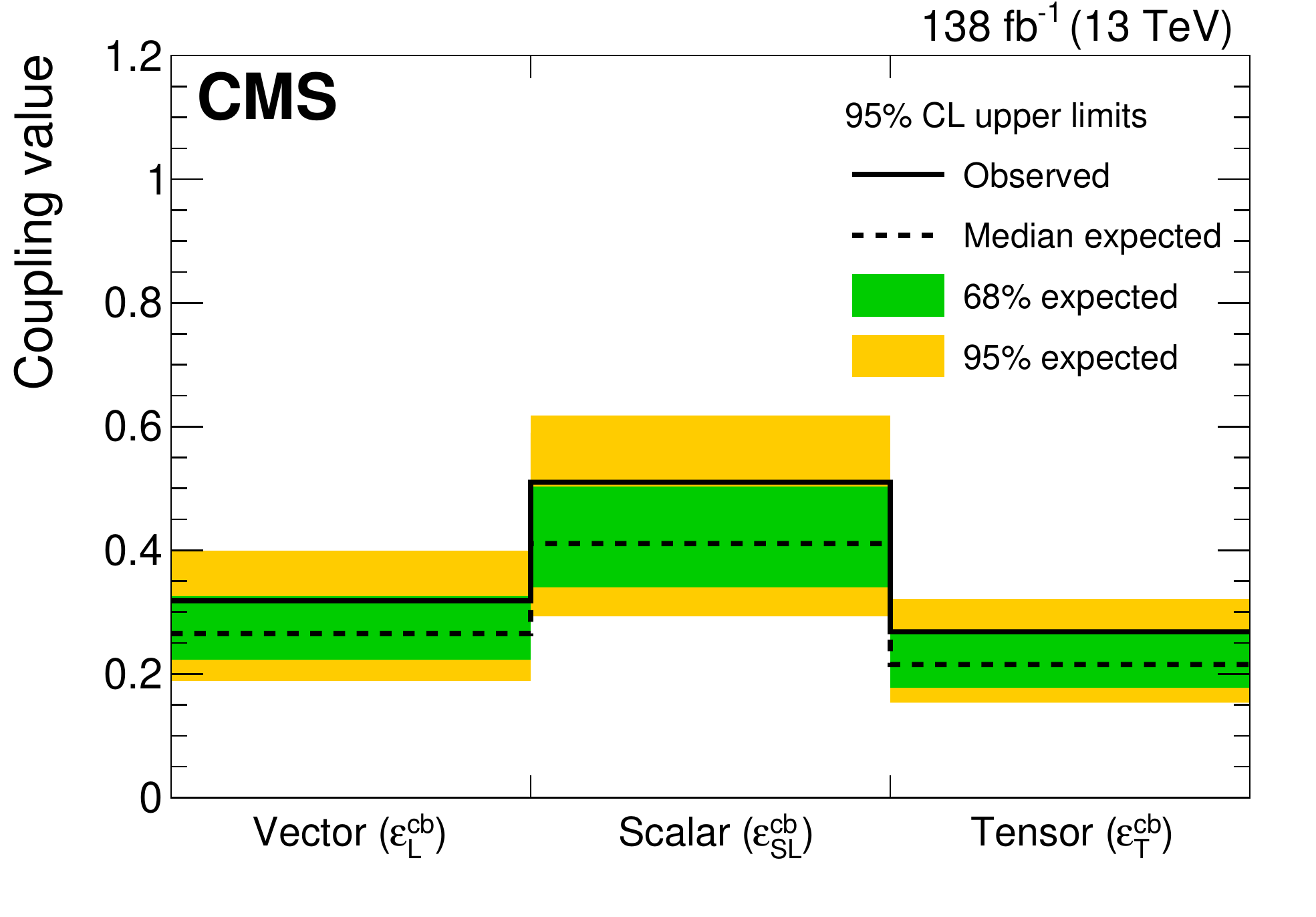}
	\caption{Bayesian upper exclusion limits at 95\% \CL on each of the Wilson coefficients 
described by the EFT model in Section~\ref{sec:Signals} based on 2016--2018 data.  
The three different coupling types represent a LH vector coupling ($\epsilon^{\text{cb}}_{\text{L}}$), scalar-tensor-like coupling 
($\epsilon^{\text{cb}}_{\text{S}_{\text{L}}}$, and tensor-like 
coupling ($\epsilon^{\text{cb}}_{\text{T}}$)). 
The 68 and 95\% quantiles of the limits are represented by the green and yellow bands, respectively.}
	\label{fig:EFTLimit}
\end{figure}

In the EFT description, we calculate limits in terms of the three Wilson 
coefficients used in this model. For each available signal hypothesis, the event yield is scaled with the square of the 
corresponding Wilson coefficient to probe a range between 0.05 and 2.0 for each coupling type, and the direct limit on the 
maximum allowed coupling value is extracted.
The results are shown in Fig.~\ref{fig:EFTLimit}.
The excluded values are 0.32 (0.27) / 0.51 (0.41) / 0.27 (0.22) observed (expected) for $\epsilon_{\text{L}}$ / 
$\epsilon_{\text{SL}}$ / 
$\epsilon_{\text{T}}$, respectively. The small difference originates from the shape of the signals, for which the 
\mT-region between 1.0 and 1.5\TeV has the most impact on the exclusion limits~\cite{Greljo:2018tzh}.

\subsection{Overview of the exclusion limits}

\begin{table}[hbt]
    \centering
    \topcaption{Summary of 95\% \CL exclusion limits (expected and observed) derived from 2016--2018 data, 
    for the physics models studied in this analysis: sequential standard model (SSM), nonuniversal gauge interaction
    model (NUGIM), a quantum black hole (QBH) interpretation, $t$-channel leptoquark (LQ), 
    and effective field interpretation (EFT).
    }
    \begin{tabular}{lccc}
    Model     & Parameter  & Expected Limit & Observed Limit \\ 
    \hline
    SSM $\PWpr\to\PGt+\PGn$     & $m_{\PWpr}$     & 4.8\TeV  & 4.8\TeV   \\
    NUGIM cot$(\theta_E)$=1     & $m_{\PWpr}$     & 4.8\TeV  & 4.8\TeV \\
    NUGIM cot$(\theta_E)$=5.5   & $m_{\PWpr}$     & 2.4\TeV  & 2.2\TeV \\
    QBH                         & $m_{\text{QBH}}$         & 6.6\TeV  & 6.6\TeV \\
    LQ democratic, $g_{\text{U}}$=1.0     & $m_{\text{LQ}}$          & 6.7\TeV  & 5.9\TeV \\
    LQ best-fit LH, $g_{\text{U}}$=1.0    & $m_{\text{LQ}}$          & 145\GeV  & 205\GeV \\
    LQ best-fit LH, $g_{\text{U}}$=2.5    & $m_{\text{LQ}}$          & 1.8\TeV  & 1.5\TeV \\
    LQ best-fit LH+RH, $g_{\text{U}}$=1.0 & $m_{\text{LQ}}$          & 645\GeV  & 515\GeV \\
    LQ best-fit LH+RH, $g_{\text{U}}$=2.5 & $m_{\text{LQ}}$          & 3.0\TeV  & 2.5\TeV \\
    EFT             & $\epsilon^{\text{cb}}_{\text{L}}$   & 0.27      & 0.32 \\ 
    EFT             & $\epsilon^{\text{cb}}_{\text{SL}}$  & 0.41      & 0.51 \\
    EFT             & $\epsilon^{\text{cb}}_{\text{T}}$   & 0.22      & 0.27 \\
    \end{tabular}
    \label{tab:limits}
\end{table}

A summary of mass exclusion limits calculated at 95\% \CL is shown in Table~\ref{tab:limits}.
The limits presented in this analysis are the most stringent to date for all the models considered.

\clearpage

\section{Summary}
\label{sec:summary}

A search for new resonant phenomena in the transverse mass distribution of a hadronically decaying 
\PGt lepton and missing transverse momentum final state has been performed.    
The analysis uses 2016--2018 data collected with the CMS detector in proton-proton collisions  with 
$\sqrt{s}=13$\TeV at the LHC, corresponding to an integrated luminosity of 138~\fbinv.    
The newly developed \textsc{DeepTau} algorithm is used to identify \PGt lepton candidates and 
to separate signal from 
background.    
Background from jets being misidentified as hadronically decaying \PGt lepton candidates is derived from 
data.    

No significant deviation from the standard model expectations is observed.
Upper limits are set at 95\% confidence level on the product of the production cross section and branching fraction 
$\sigma\mathcal{B}$ of a 
new resonance gauge boson (\PWpr) decaying to a \PGt lepton and a neutrino.   
Lower limits are derived on the mass of a \PWpr boson  in the sequential standard model 
and on the mass of a quantum black hole. 
In the case of the generalized coupling model,
upper limits on $\sigma\mathcal{B}$ are translated into upper exclusion limits on the ratio of 
the couplings of the \PWpr boson to that of the standard model \PW boson, $\WpCoupling/\Wcoupling$.
Similarly, limits are obtained on
the mixing angle $\theta_{E}$ for the nonuniversal gauge interaction model
and on the Wilson coefficients for an effective field theory model.

The process of $t$-channel leptoquark (LQ) exchange is targeted explicitly in this analysis. 
For the first time, upper limits are placed 
on the cross section of the $\Pp{}\Pp\to\PGt\PGn$ process mediated by $t$-channel LQ exchange.
Three benchmark coupling scenarios are tested 
that correspond to the best fits to the flavor anomaly data and to a generic democratic flavor structure. 
Significant portions of the LQ  
parameter space are excluded by placing upper limits on the LQ coupling $g_{\text{U}}$ as a function of the LQ mass, 
including parts of the 
region preferred by a vector LQ explanation of the anomalies. Additionally, a model-independent limit is provided. The limits obtained for the various interpretations presented are the most stringent to date.

\begin{acknowledgments}
We congratulate our colleagues in the CERN accelerator departments for the excellent performance of the LHC and thank the technical and administrative staffs at CERN and at other CMS institutes for their contributions to the success of the CMS effort. In addition, we gratefully acknowledge the computing centers and personnel of the Worldwide LHC Computing Grid and other centers for delivering so effectively the computing infrastructure essential to our analyses. Finally, we acknowledge the enduring support for the construction and operation of the LHC, the CMS detector, and the supporting computing infrastructure provided by the following funding agencies: BMBWF and FWF (Austria); FNRS and FWO (Belgium); CNPq, CAPES, FAPERJ, FAPERGS, and FAPESP (Brazil); MES and BNSF (Bulgaria); CERN; CAS, MoST, and NSFC (China); MINCIENCIAS (Colombia); MSES and CSF (Croatia); RIF (Cyprus); SENESCYT (Ecuador); MoER, ERC PUT and ERDF (Estonia); Academy of Finland, MEC, and HIP (Finland); CEA and CNRS/IN2P3 (France); BMBF, DFG, and HGF (Germany); GSRI (Greece); NKFIH (Hungary); DAE and DST (India); IPM (Iran); SFI (Ireland); INFN (Italy); MSIP and NRF (Republic of Korea); MES (Latvia); LAS (Lithuania); MOE and UM (Malaysia); BUAP, CINVESTAV, CONACYT, LNS, SEP, and UASLP-FAI (Mexico); MOS (Montenegro); MBIE (New Zealand); PAEC (Pakistan); MES and NSC (Poland); FCT (Portugal); MESTD (Serbia); MCIN/AEI and PCTI (Spain); MOSTR (Sri Lanka); Swiss Funding Agencies (Switzerland); MST (Taipei); MHESI and NSTDA (Thailand); TUBITAK and TENMAK (Turkey); NASU (Ukraine); STFC (United Kingdom); DOE and NSF (USA).

\hyphenation{Rachada-pisek} Individuals have received support from the Marie-Curie program and the European Research Council and Horizon 2020 Grant, contract Nos.\ 675440, 724704, 752730, 758316, 765710, 824093, 884104, and COST Action CA16108 (European Union); the Leventis Foundation; the Alfred P.\ Sloan Foundation; the Alexander von Humboldt Foundation; the Belgian Federal Science Policy Office; the Fonds pour la Formation \`a la Recherche dans l'Industrie et dans l'Agriculture (FRIA-Belgium); the Agentschap voor Innovatie door Wetenschap en Technologie (IWT-Belgium); the F.R.S.-FNRS and FWO (Belgium) under the ``Excellence of Science -- EOS" -- be.h project n.\ 30820817; the Beijing Municipal Science \& Technology Commission, No. Z191100007219010; the Ministry of Education, Youth and Sports (MEYS) of the Czech Republic; the Hellenic Foundation for Research and Innovation (HFRI), Project Number 2288 (Greece); the Deutsche Forschungsgemeinschaft (DFG), under Germany's Excellence Strategy -- EXC 2121 ``Quantum Universe" -- 390833306, and under project number 400140256 - GRK2497; the Hungarian Academy of Sciences, the New National Excellence Program - \'UNKP, the NKFIH research grants K 124845, K 124850, K 128713, K 128786, K 129058, K 131991, K 133046, K 138136, K 143460, K 143477, 2020-2.2.1-ED-2021-00181, and TKP2021-NKTA-64 (Hungary); the Council of Science and Industrial Research, India; the Latvian Council of Science; the Ministry of Education and Science, project no. 2022/WK/14, and the National Science Center, contracts Opus 2021/41/B/ST2/01369 and 2021/43/B/ST2/01552 (Poland); the Funda\c{c}\~ao para a Ci\^encia e a Tecnologia, grant CEECIND/01334/2018 (Portugal); the National Priorities Research Program by Qatar National Research Fund; MCIN/AEI/10.13039/501100011033, ERDF ``a way of making Europe", and the Programa Estatal de Fomento de la Investigaci{\'o}n Cient{\'i}fica y T{\'e}cnica de Excelencia Mar\'{\i}a de Maeztu, grant MDM-2017-0765 and Programa Severo Ochoa del Principado de Asturias (Spain); the Chulalongkorn Academic into Its 2nd Century Project Advancement Project, and the National Science, Research and Innovation Fund via the Program Management Unit for Human Resources \& Institutional Development, Research and Innovation, grant B05F650021 (Thailand); the Kavli Foundation; the Nvidia Corporation; the SuperMicro Corporation; the Welch Foundation, contract C-1845; and the Weston Havens Foundation (USA).
  
\end{acknowledgments}

\bibliography{auto_generated} 

\cleardoublepage \appendix\section{The CMS Collaboration \label{app:collab}}\begin{sloppypar}\hyphenpenalty=5000\widowpenalty=500\clubpenalty=5000
\cmsinstitute{Yerevan Physics Institute, Yerevan, Armenia}
{\tolerance=6000
A.~Tumasyan\cmsAuthorMark{1}\cmsorcid{0009-0000-0684-6742}
\par}
\cmsinstitute{Institut f\"{u}r Hochenergiephysik, Vienna, Austria}
{\tolerance=6000
W.~Adam\cmsorcid{0000-0001-9099-4341}, J.W.~Andrejkovic, T.~Bergauer\cmsorcid{0000-0002-5786-0293}, S.~Chatterjee\cmsorcid{0000-0003-2660-0349}, K.~Damanakis\cmsorcid{0000-0001-5389-2872}, M.~Dragicevic\cmsorcid{0000-0003-1967-6783}, A.~Escalante~Del~Valle\cmsorcid{0000-0002-9702-6359}, P.S.~Hussain\cmsorcid{0000-0002-4825-5278}, M.~Jeitler\cmsAuthorMark{2}\cmsorcid{0000-0002-5141-9560}, N.~Krammer\cmsorcid{0000-0002-0548-0985}, L.~Lechner\cmsorcid{0000-0002-3065-1141}, D.~Liko\cmsorcid{0000-0002-3380-473X}, I.~Mikulec\cmsorcid{0000-0003-0385-2746}, P.~Paulitsch, F.M.~Pitters, J.~Schieck\cmsAuthorMark{2}\cmsorcid{0000-0002-1058-8093}, R.~Sch\"{o}fbeck\cmsorcid{0000-0002-2332-8784}, D.~Schwarz\cmsorcid{0000-0002-3821-7331}, M.~Sonawane\cmsorcid{0000-0003-0510-7010}, S.~Templ\cmsorcid{0000-0003-3137-5692}, W.~Waltenberger\cmsorcid{0000-0002-6215-7228}, C.-E.~Wulz\cmsAuthorMark{2}\cmsorcid{0000-0001-9226-5812}
% ...
\par}
\vskip\cmsinstskip
\dag:~Deceased\\
$^{1}$Also at Yerevan State University, Yerevan, Armenia\\
$^{2}$Also at TU Wien, Vienna, Austria\\
% ...
\end{sloppypar}
%%% END EDITABLE REGION %%%
% skeleton_end
\end{document}